\documentclass[12pt]{article}
\newcommand{\nc}{\newcommand}
\nc{\renc}{\renewcommand}

\usepackage{calc}
\usepackage{ifthen}
\usepackage{graphicx}
\usepackage{epsfig}
\usepackage{miniplot}

\usepackage{wrapfig}
\usepackage{subfigure}
\usepackage{rotfloat}

\usepackage{epsfig}
\usepackage{graphicx}
\usepackage{epsf}

\nc{\half}{{\textstyle{1\over2}}}
\nc{\etal}{\mbox{\it et al. }}
\nc{\ie}{{\it i.e.}}
\nc{\eg}{{\it e.g.}}

\renc{\thefootnote}{\arabic{footnote}}
\nc{\capt}[1]{{\bf Figure.} {\small\sl #1}}

\nc{\eqs}[2]{\mbox{Eqs.~(\ref{#1},\,\ref{#2})}}
\nc{\eq}[1]{\mbox{Eq.~(\ref{#1})}}

\nc{\figs}[2]{\mbox{Figs.~(\ref{#1},\,\ref{#2})}}
\nc{\fig}[1]{\mbox{Fig~.(\ref{#1})}}

\nc{\tag}[1]{\label{#1} \marginpar{{\footnotesize #1}}}
\nc{\mtag}[1]{\label{#1} \mbox{\marginpar{{\footnotesize #1}}}}
\renc{\baselinestretch}{1.5}
\newlength{\overeqskip}
\newlength{\undereqskip}
\nc{\be}[1]{\begin{equation} \mbox{$\label{#1}$}}
\nc{\bea}[1]{\begin{eqnarray} \mbox{$\label{#1}$}}
\nc{\Section}[2]{\section{#2}\label{#1}}
\nc{\Bibitem}[1]{\bibitem{#1}}
\nc{\Label}[1]{\label{#1}}

\nc{\eea}{\vspace{\undereqskip}\end{eqnarray}}
\nc{\ee}{\vspace{\undereqskip}\end{equation}}
\nc{\bdm}{\begin{displaymath}}
\nc{\edm}{\end{displaymath}}
\nc{\dpsty}{\displaystyle}
\nc{\bc}{\begin{center}}
\nc{\ec}{\end{center}}
\nc{\ba}{\begin{array}}
\nc{\ea}{\end{array}}
\nc{\bab}{\begin{abstract}}
\nc{\eab}{\end{abstract}}
\nc{\btab}{\begin{tabular}}
\nc{\etab}{\end{tabular}}
\nc{\bit}{\begin{itemize}}
\nc{\eit}{\end{itemize}}
\nc{\ben}{\begin{enumerate}}
\nc{\een}{\end{enumerate}}
\nc{\bfig}{\begin{figure}}
\nc{\efig}{\end{figure}}
\nc{\arreq}{&\!=\!&}
\nc{\arrmi}{&\!-\!&}
\nc{\arrpl}{&\!+\!&}
\nc{\arrap}{&\!\!\!\approx\!\!\!&}
\nc{\non}{\nonumber\\*}
\nc{\align}{\!\!\!\!\!\!\!\!&&}

\def\lsim{\; \raise0.3ex\hbox{$<$\kern-0.75em
      \raise-1.1ex\hbox{$\sim$}}\; }
\def\gsim{\; \raise0.3ex\hbox{$>$\kern-0.75em
      \raise-1.1ex\hbox{$\sim$}}\; }
\nc{\DOT}{\hspace{-0.08in}{\bf .}\hspace{0.1in}}
\nc{\Laada}{\hbox {$\sqcap$ \kern -1em $\sqcup$}}
\nc\loota{{\scriptstyle\sqcap\kern-0.55em\hbox{$\scriptstyle\sqcup$}}}
\nc\Loota{{\sqcap\kern-0.65em\hbox{$\sqcup$}}}
\nc\laada{\Loota}
\nc{\qed}{\hskip 3em \hbox{\BOX} \vskip 2ex}

\nc{\real}{{\rm I \! R}}
\nc{\Z}{{\sf Z \!\!\! Z}}
\nc{\complex}{{\rm C\!\!\! {\sf I}\,\,}}
\def\bigid{\leavevmode\hbox{\small1\kern-3.8pt\normalsize1}}
\def\id{\leavevmode\hbox{\small1\kern-3.3pt\normalsize1}}
\nc{\slask}{\!\!\!/}
\nc{\bis}{{\prime\prime}}
\nc{\pa}{\partial}
\nc{\na}{\nabla}
\nc{\ra}{\rangle}
\nc{\la}{\langle}
\nc{\goto}{\rightarrow}
\nc{\swap}{\leftrightarrow}

\nc{\EE}[1]{ \mbox{$\cdot10^{#1}$} }
\nc{\abs}[1]{\left|#1\right|}
\nc{\at}[2]{\left.#1\right|_{#2}}
\nc{\norm}[1]{\|#1\|}
\nc{\abscut}[2]{\Abs{#1}_{\scriptscriptstyle#2}}
\nc{\vek}[1]{{\rm\bf #1}}
\nc{\integral}[2]{\int\limits_{#1}^{#2}}
\nc{\inv}[1]{\frac{1}{#1}}
\nc{\dd}[2]{{{\partial #1}\over{\partial #2}}}
\nc{\ddd}[2]{{{{\partial}^2 #1}\over{\partial {#2}^2}}}
\nc{\dddd}[3]{{{{\partial}^2 #1}\over
        {\partial #2 \partial #3}}}
\nc{\dder}[2]{{{d #1}\over{d #2}}}
\nc{\ddder}[2]{{{d^2 #1}\over{d {#2}^2}}}
\nc{\dddder}[3]{{d^2 #1}\over
        {d #2 d #3}}
\nc{\dx}[1]{d\,^{#1}x}
\nc{\dy}[1]{d\,^{#1}y}
\nc{\dz}[1]{d\,^{#1}z}
\nc{\dl}[1]{\frac{d\,^{#1}l}{(2\pi)^{#1}}}
\nc{\dk}[1]{\frac{d\,^{#1}k}{(2\pi)^{#1}}}
\nc{\dq}[1]{\frac{d\,^{#1}q}{(2\pi)^{#1}}}

\nc{\cc}{\mbox{$c.c.$ }}
\nc{\hc}{\mbox{$h.c.$ }}
\nc{\cf}{cf.\ }
\nc{\erfc}{{\rm erfc}}
\nc{\Tr}{{\rm Tr\,}}
\nc{\tr}{{\rm tr\,}}
\nc{\pol}{{\rm pol}}
\nc{\sign}{{\rm sign}}
\nc{\bfT}{{\bf T }}

\def\GeV{{\rm\ GeV}}

\nc{\cA}{{\cal A}}
\nc{\cB}{{\cal B}}
\nc{\cD}{{\cal D}}
\nc{\cE}{{\cal E}}
\nc{\cG}{{\cal G}}
\nc{\cH}{{\cal H}}
\nc{\cL}{{\cal L}}
\nc{\cO}{{\cal O}}
\nc{\cT}{{\cal T}}
\nc{\cN}{{\cal N}}
\nc{\rvac}[1]{|{\cal O}#1\rangle}
\nc{\lvac}[1]{\langle{\cal O}#1|}
\nc{\rvacb}[1]{|{\cal O}_\beta #1\rangle}
\nc{\lvacb}[1]{\langle{\cal O}_\beta #1 |}
\nc{\bb}{\bar{\beta}}
\nc{\bt}{\tilde{\beta}}
\nc{\ctH}{\tilde{\cal H}}
\nc{\chH}{\hat{\cal H}}
\nc{\1}{\aa}
\nc{\2}{\"{a}}
\nc{\3}{\"{o}}
\nc{\4}{\AA}
\nc{\5}{\"{A}}
\nc{\6}{\"{O}}
\nc{\al}{\alpha}
\nc{\g}{\gamma}
\nc{\Del}{\Delta}
\nc{\e}{\epsilon}
\nc{\eps}{\epsilon}
\nc{\lam}{\lambda}
\nc{\om}{\omega}
\nc{\Om}{\Omega}
\nc{\ve}{\varepsilon}
\nc{\mn}{{\mu\nu}}
\nc{\vp}{\varphi}

\nc{\advp}[3]{{\it  Adv.\ in\ Phys.\ }{{\bf #1} {(#2)} {#3}}}
\nc{\annp}[3]{{\it  Ann.\ Phys.\ (N.Y.)\ }{{\bf #1} {(#2)} {#3}}}
\nc{\apl}[3]{{\it  Appl. Phys. Lett. }{{\bf #1} {(#2)} {#3}}}
\nc{\apj}[3]{{\it  Ap.\ J.\ }{{\bf #1} {(#2)} {#3}}}
\nc{\apjl}[3]{{\it  Ap.\ J.\ Lett.\ }{{\bf #1} {(#2)} {#3}}}
\nc{\app}[3]{{\it Astropart.\ Phys.\ }{{\bf #1} {(#2)} {#3}}}
\nc{\cmp}[3]{{\it  Comm.\ Math.\ Phys.\ }{{ \bf #1} {(#2)} {#3}}}
\nc{\cqg}[3]{{\it  Class.\ Quant.\ Grav.\ }{{\bf #1} {(#2)} {#3}}}
\nc{\epl}[3]{{\it  Europhys.\ Lett.\ }{{\bf #1} {(#2)} {#3}}}
\nc{\ijmp}[3]{{\it Int.\ J.\ Mod.\ Phys.\ }{{\bf #1} {(#2)} {#3}}}
\nc{\ijtp}[3]{{\it Int.\ J.\ Theor.\ Phys.\ }{{\bf #1} {(#2)} {#3}}}
\nc{\jmp}[3]{{\it  J.\ Math.\ Phys.\ }{{ \bf #1} {(#2)} {#3}}}
\nc{\jpa}[3]{{\it  J.\ Phys.\ A\ }{{\bf #1} {(#2)} {#3}}}
\nc{\jpc}[3]{{\it  J.\ Phys.\ C\ }{{\bf #1} {(#2)} {#3}}}
\nc{\jap}[3]{{\it J.\ Appl.\ Phys.\ }{{\bf #1} {(#2)} {#3}}}
\nc{\jpsj}[3]{{\it J.\ Phys.\ Soc.\ Japan\ }{{\bf #1} {(#2)} {#3}}}
\nc{\lmp}[3]{{\it Lett.\ Math.\ Phys.\ }{{\bf #1} {(#2)} {#3}}}
\nc{\mpl}[3]{{\it  Mod.\ Phys.\ Lett.\ }{{\bf #1} {(#2)} {#3}}}
\nc{\ncim}[3]{{\it  Nuov.\ Cim.\ }{{\bf #1} {(#2)} {#3}}}
\nc{\np}[3]{{\it  Nucl.\ Phys.\ }{{\bf #1} {(#2)} {#3}}}
\nc{\npps}[3]{{\it  Nucl.\ Phys.\ Proc.\ Suppl.\ }{{\bf #1} {(#2)} {#3}}}
\nc{\pr}[3]{{\it Phys.\ Rev.\ }{{\bf #1} {(#2)} {#3}}}
\nc{\pra}[3]{{\it  Phys.\ Rev.\ A\ }{{\bf #1} {(#2)} {#3}}}
\nc{\prb}[3]{{\it  Phys.\ Rev.\ B\ }{{{\bf #1} {(#2)} {#3}}}}
\nc{\prc}[3]{{\it  Phys.\ Rev.\ C\ }{{\bf #1} {(#2)} {#3}}}
\nc{\prd}[3]{{\it  Phys.\ Rev.\ D\ }{{\bf #1} {(#2)} {#3}}}
\nc{\prl}[3]{{\it Phys.\ Rev.\ Lett.\ }{{\bf #1} {(#2)} {#3}}}
\nc{\pl}[3]{{\it  Phys.\ Lett.\ }{{\bf #1} {(#2)} {#3}}}
\nc{\prep}[3]{{\it Phys.\ Rep.\ }{{\bf #1} {(#2)} {#3}}}
\nc{\prsl}[3]{{\it Proc.\ R.\ Soc.\ London\ }{{\bf #1} {(#2)} {#3}}}
\nc{\ptp}[3]{{\it  Prog.\ Theor.\ Phys.\ }{{\bf #1} {(#2)} {#3}}}
\nc{\ptps}[3]{{\it  Prog\ Theor.\ Phys.\ suppl.\ }{{\bf #1} {(#2)} {#3}}}
\nc{\physa}[3]{{\it  Physica\ A\ }{{\bf #1} {(#2)} {#3}}}
\nc{\physb}[3]{{\it  Physica\ B\ }{{\bf #1} {(#2)} {#3}}}
\nc{\phys}[3]{{\it Physica\ }{{\bf #1} {(#2)} {#3}}}
\nc{\rmp}[3]{{\it  Rev.\ Mod.\ Phys.\ }{{\bf #1} {(#2)} {#3}}}
\nc{\rpp}[3]{{\it Rep.\ Prog.\ Phys.\ }{{\bf #1} {(#2)} {#3}}}
\nc{\sjnp}[3]{{\it Sov.\ J.\ Nucl.\ Phys.\ }{{\bf #1} {(#2)} {#3}}}
\nc{\spjetp}[3]{{\it Sov.\ Phys.\ JETP\ }{{\bf #1} {(#2)} {#3}}}
\nc{\yf}[3]{{\it Yad.\ Fiz.\ }{{\bf #1} {(#2)} {#3}}}
\nc{\zetp}[3]{{\it Zh.\ Eksp.\ Teor.\ Fiz.\  }{{\bf #1}  {(#2)} {#3}}}
\nc{\zp}[3]{{\it Z.\ Phys.\ }{{\bf #1} {(#2)} {#3}}}
\nc{\ibid}[3]{{\sl ibid.\ }{{\bf #1} {#2} {#3}}}
\nc{\rf}[1]{(\ref{#1})}
\nc{\nn}{\nonumber \\*}
\nc{\bfB}{\bf{B}}
\nc{\bfv}{\bf{v}}
\nc{\bfx}{\bf{x}}
\nc{\bfy}{\bf{y}}
\nc{\vx}{\vec{x}}
\nc{\vy}{\vec{y}}
\nc{\oB}{\overline{B}}
\nc{\oI}{\overline{I}}
\nc{\oR}{\overline{R}}
\nc{\rar}{\rightarrow}
\nc{\ti}{\times}
\nc{\slsh}{\hskip-5pt/}
\nc{\sm}{Standard~Model~}
\nc{\MP}{M_{\rm Pl}}
\nc{\tp}{t_{\rm Pl}}
\nc{\ave}{\bar{E}}

\nc{\eff}{{\rm eff}}
\nc{\kk}{\vek{k}}
\nc{\pp}{{\rm p}}
\nc{\ga}{g_{a\gamma}}
\nc{\vv}{\\}
\nc{\eee}{{\bf E}}
\nc{\bbb}{{\bf B}}
\nc{\qcd}{T_{\rm QCD}}
\nc{\G}{\rm \ G}
\def\vec#1{{\bf #1}}

\def\lae{\;^{<}_{\sim} \;} \def\gae{\; ^{>}_{\sim} \;} 

\def\ell{e^{c}LL}

\begin{document}
{\title{\vskip-2truecm{\hfill {{\small \\
	\hfill \\
	}}\vskip 1truecm}
{\LARGE  Gauge Singlet Scalars as Cold Dark Matter
}}
{\author{
{\sc \large John McDonald$^{1}$}\\
{\sl\small Cosmology and Astroparticle Physics Group, University of Lancaster,
Lancaster LA1 4YB, UK}
}
\maketitle
\begin{abstract}
\noindent

                In light of recent interest in minimal extensions of the Standard Model and gauge singlet scalar cold dark matter, we provide an arXiv preprint of the paper, published as Phys.Rev. D50 (1994) 3637, which presented the first detailed analysis of gauge singlet scalar cold dark matter.

\end{abstract} 
\vfil
 \footnoterule{\small   $^1$j.mcdonald@lancaster.ac.uk}   
 \newpage 
\setcounter{page}{1}

\begin{center} {\bf \large Abstract} \end{center}  

      We consider a very simple extension of the standard
model in which one or more gauge singlet scalars $\rm S_{i}$
couples to the standard model via an interaction of the form 
$\rm \lambda_{S}S^{\dagger}_{i}S_{i}H^{\dagger}H$, where H
is the standard model Higgs doublet. The thermal relic
density of S scalars is calculated as a function of $\rm
\lambda_{S}$ and the S mass $\rm m_{S}$. The regions of the
$\rm (m_{S},\lambda_{S})$ parameter space which can be probed
by present and future experiments to detect scattering
of S dark matter particles from Ge nuclei, and to observe
upward-moving muons and contained events in neutrino
detectors due to high-energy neutrinos from annihilations of
S dark matter particles in the Sun and Earth, are discussed.
Present experimental bounds place only very weak constraints 
on the possibility of thermal relic S scalar dark matter.
The next generation of cryogenic Ge detectors and of large area 
($\rm 10^{4}m^{2}$) neutrino detectors will be able to
investigate most of the parameter space corresponding to thermal 
relic S scalar dark matter with $\rm m_{S} \; ^{<}_{\sim} \; 
50GeV$, while a $\rm 1 \;km^{2}$ detector would in general be
able to detect thermal relic S scalars with $\rm m_{S} \;
^{<}_{\sim} \; 100GeV$ as a dark matter candidate and would be able to detect 
up to $\rm m_{S} \; ^{<}_{\sim} \; 500GeV$ or more if the Higgs
boson is lighter than 100GeV.

\section{Introduction}

          There is strong evidence that the mass density of
the Universe is mainly composed of some non-hadronic form of
dark matter$\rm ^{[1,2]}$. Direct observation of galaxies
and clusters of galaxies$\rm ^{[2]}$ indicates that $\rm
\Omega = $ 0.1 to 0.3, where $\rm \Omega$ is the ratio of
the mass density to the critical density in the Universe at
present. Nucleosynthesis constrains the density of hadronic
dark matter to satisfy$\rm ^{[3]}$ $\rm 0.011 <
\Omega_{B}h^{2} < 0.019$, where $\rm h = 0.5-1$
parameterizes the uncertainty in the 
observed value of the Hubble parameter. Inflation and
naturalness considerations$\rm ^{[4]}$ suggest that $\rm
\Omega = 1$. Although it seems possible that baryons could
just about account for $\rm \Omega = 0.1$ dark matter, it
would not be possible for primordial density perturbations
to grow sufficiently in a baryon dominated Universe to allow 
galaxy formation$\rm ^{[5]}$ to be consistent with the
magnitude of temperature fluctuations of the cosmic
microwave background radiation as observed by COBE$\rm
^{[6]}$. This requires the addition of a density of
non-hadronic dark matter, preferably cold dark matter
(CDM)$\rm ^{[5]}$. It would also be difficult to explain, if
halo dark matter was hadronic in nature, how all the hadrons 
in galactic halos could be hidden$\rm ^{[7]}$. Searches for faint stars support 
the conclusion that the halo dark matter cannot primarily be baryonic$\rm ^{[8]}$ 
(although recent observations of microlensing by dark objects in the galactic halo 
do show that at least some baryonic halo dark matter exists$\rm ^{[9]}$). Thus it is
likely that the Universe is dominated by a density of CDM
satisfying $\rm \Omega_{CDM} \; ^{>}_{\sim} \; 0.1$. The age 
of the Universe imposes an upper limit on $\rm \Omega$, $\rm
\Omega h^{2} \; ^{<}_{\sim} \; 1$ $\rm ^{[1]}$. This leaves
a window for which a density of particles can consistently
serve as the primary component of the halo dark matter, $\rm 
\; 0.025 \; ^{<}_{\sim} \; \Omega h^{2} 
\; ^{<}_{\sim} \; 1$.    

        In this paper we will study in some detail an
extremely simple and natural extension of the $\rm SU(3)_{c} \times SU(2)_{L} \times U(1)_{Y}$ standard
model, namely, the addition of one or more gauge singlet
complex scalars $\rm S_{i}$. These
scalars, if stable, can in principle account for a density
of CDM. Stability of the scalars can most simply be guaranteed if a
continuous or discrete symmetry exists under which the gauge singlet 
scalars are the lightest particles transforming non-trivially. 
(Additional continuous and discrete symmetries are a common feature of many extensions of the standard model, serving to simplify the models and to eliminate phenomenologically unwelcome interactions such as those leading to baryon and lepton number violation or to flavor-changing neutral currents.) In addition, it is necessary that the $\rm S_{i}$ 
do not acquire vacuum expectation values, which in turn requires that they have positive mass squared terms. This model for CDM is essentially determined by just 
three parameters: the Higgs boson mass $\rm m_{h}$, the S
scalar mass $\rm m_{S}$, and the coupling of the S scalars
to the Higgs bosons $\rm \lambda_{S}$.
In particular, we will consider the thermal relic density of
S scalars, coming from S scalars freezing-out of thermal 
equilibrium. This is the simplest and most natural origin of 
a relic density of S scalars, although in principle other
possibilities exist, such as S scalars originating in the
out-of-equilibrium decay of some heavy particle. We will be
particularly interested in the possibility of detecting S
scalar cold dark matter, as a function of $\rm \lambda_{S}$,
$\rm m_{S}$ and $\rm m_{h}$, either via direct detection of the recoil
energy coming from elastic scattering of S dark matter
particles from Ge nuclei$\rm ^{[10,11]}$, or by observing
upward moving muons or contained events in neutrino
detectors, produced by high-energy neutrinos coming from S
annihilations in the Sun or in the Earth$\rm
^{[12-17]}$.

              The paper is organized as follows. In section
2 we discuss the thermal relic density of gauge singlet
scalars in the Universe at present. In section 3 we discuss 
the elastic scattering of S scalars from Ge nuclei. In
section 4 we discuss the rate of upward moving muons and
contained events produced by high energy neutrinos 
due to S annihilations in the core of the Sun and of the
Earth. In section 5 we give our conclusions. In the Appendix 
we give some details of the calculation of the upward-moving 
muon and contained event rates.  

\section{S scalar dark matter}    

          We consider extending the standard model by the
addition of terms involving the S scalars
$${\rm L_{S} = \partial^{\mu}S_{i}^{\dagger}
\partial_{\mu}S_{i} -m^{2} S_{i}^{\dagger}S_{i} -
\lambda_{S}S_{i}^{\dagger}S_{i}H^{\dagger}H   }\eqno(2.1),$$
where $i = 1,...,N$. This model  has a global U(1) symmetry, 
$\rm S_{i} \rightarrow e^{i \alpha} S_{i}$, which guarantees
the stability of the $\rm S_{i}$ scalars by eliminating the
interaction terms involving odd powers of $\rm S_{i}$
and $\rm S_{i}^{\dagger}$ which lead to $\rm S_{i}$ decay.
We first consider the case $\rm N =1$. 
In order to calculate the relic density
arising from S scalars freezing out of thermal equilibrium
we will use the usual Lee-Weinberg (LW) approximation$\rm
^{[18]}$ to solve the rate equation for the density of $\rm
S_{i}$ scalars. The rate
equation is given by
$${\rm  \frac{dn_{S}}{dt} = -3Hn_{S} -
<\sigma_{ann} v_{rel}> (n_{S}^{2}-n_{o}^{2}) }\eqno(2.2).$$
$\rm \sigma_{ann}$ is the $\rm S S^{\dagger}$ annihilation
cross-section, $\rm v_{rel}$ is the relative velocity of
the annihilating particles and $H$ is the expansion rate of
the Universe. The angular brackets denote the thermal average value. 
(2.2) gives the number density of S scalars $\rm n_{S}$. The 
total density of $\rm S$ and $\rm S^{\dagger}$ scalars is
then $\rm 2 n_{S}$. The equilibrium S density $\rm n_{o}$,
for $\rm m_{S}/T \gg 1$, is given by 
$${\rm n_{o} = T^{3} \left(\frac{m_{S}}{2\pi T}\right)^{3/2} 
e^{-\frac{m_{S}}{T}}   }\eqno(2.3).$$
The approximate solution of (2.2) is then found by 
rewriting (2.2) as
$${\rm \frac{df}{dT} = \frac{<\sigma_{ann}
v_{rel}>}{K}\left(f^{2} - f_{o}^{2}\right)   }\eqno(2.4),$$
where $\rm f = \frac{n_{S}}{T^{3}}$, $\rm f_{o} =
\frac{n_{o}}{T^{3}}$ and $\rm K = \left(4 \pi^{3}
\overline{g}(T)/45 M_{Pl}^{2} \right)^{1/2}$. $\rm
\overline{g}(T)$ is the number of degrees of freedom with
masses smaller than T. In this we are assuming that the
number of degrees of freedom in thermal equilibrium with the 
photons, $g(T)$, is constant around the S freeze-out
temperature $\rm T_{fS}$, i.e. no particle thresholds at
$\rm T \approx T_{fS}$, and also that the Universe is
radiation dominated. The LW solution is given by assuming
that $\rm f=f_{o}$ until the temperature at which
$${\rm \left|\frac{df_{o}}{dT}\right| \; = \;
\frac{<\sigma_{ann} v_{rel}>}{K} \; f_{o}^{2} }\eqno(2.5)$$
is satisfied, which defines the S freeze-out temperature,
$\rm T_{fS}$.
Then for $\rm T < T_{fS}$ one solves (2.4) with $\rm
f_{o}=0$ on the right hand side and with $\rm f(T_{fS}) =
f_{o}(T_{fS})$. The freeze-out temperature is then obtained
from
$${\rm x_{fS}^{-1} = \ln \left(\frac{m_{S} x_{fS}^{2}
A}{\left(1-3 x_{fS}/2\right)\left(2 \pi
x_{fS}\right)^{3/2} } \right)    }\eqno(2.6),$$
where $\rm x_{fS} = T_{fS}/m_{S} $ and $\rm A =
\frac{<\sigma_{ann} v_{rel}> }{K} $. The present total mass
density in S scalars and antiscalars is then 
$${\rm \Omega_{S} \equiv
\frac{\rho_{S}+\rho_{S^{\dagger}}}{\rho_{c}} = 
2\frac{g(T_{\gamma})}{g(T_{fS})} \frac{K}{T_{\gamma}
x_{fS} 
<\sigma_{ann} v_{rel}>
}\left(\frac{T_{\gamma}^{4}}{\rho_{c}}\right)
\frac{\left(1-3x_{fS}/2\right)}{\left(1-x_{fS}/2\right)}    
}\eqno(2.7),$$
where it has been assumed that $\rm <\sigma_{ann} v_{rel}>$
is T independent. $\rm \rho_{c} = 7.5x10^{-47}h^{2}GeV^{4}$
is the critical closure density of the Universe at present
(h = 0.5 to 1) and $\rm T_{\gamma}$ is the present photon
temperature.

        In order to calculate $\rm <\sigma_{ann}
v_{rel}>$ we need the $\rm S S^{\dagger}$ annihilation
modes. These are shown in Figure 1. The corresponding
contributions to $\rm <\sigma_{ann} v_{rel}>$ are given by
\newline $\rm \underline{SS^{\dagger} \rightarrow
h^{o}h^{o}}:$
$${\rm  \frac{\lambda_{S}^{2}}{64 \pi m_{S}^{2}}
\left(1-\frac{m_{h}^{2}}{m_{S}^{2}}\right)^{1/2}
}\eqno(2.8a),$$ 
\newline $\rm \underline{SS^{\dagger} \rightarrow W^{+}W^{-} 
}:$
$${\rm  2\left(1+\frac{1}{2}
\left(1-\frac{2 m_{S}^{2}}{m_{W}^{2}}\right)^{2}\right)
\frac{\lambda_{S}^{2} m_{W}^{4}}{8 \pi m_{S}^{2} 
\left(\left(4 m_{S}^{2}
-m_{h}^{2}\right)^{2}+m_{h}^{2}\Gamma_{h}^{2}\right)  }
\left(1-\frac{m_{W}^{2}}{m_{S}^{2}}\right)^{1/2}   
}\eqno(2.8b),$$
\newline $\rm \underline{SS^{\dagger} \rightarrow Z^{o}Z^{o} 
}:$
$${\rm  2\left(1+\frac{1}{2}
\left(1-\frac{2 m_{S}^{2}}{m_{Z}^{2}}\right)^{2}\right)
\frac{\lambda_{S}^{2} m_{Z}^{4}}{16 \pi m_{S}^{2} 
\left(\left(4 m_{S}^{2}
-m_{h}^{2}\right)^{2}+m_{h}^{2}\Gamma_{h}^{2}\right)  }
\left(1-\frac{m_{Z}^{2}}{m_{S}^{2}}\right)^{1/2}
}\eqno(2.8c),$$
\newline $\rm \underline{SS^{\dagger} \rightarrow
\overline{f}f}:$
$${\rm  \frac{m_{W}^{2}}{\pi g^{2}} \frac{\lambda_{f}^{2}
\lambda_{S}^{2} }{ 
\left(\left(4m_{S}^{2}-m_{h}^{2}\right)^{2}
+m_{h}^{2}\Gamma_{h}^{2}\right) }
\left(1-\frac{m_{f}^{2}}{m_{S}^{2}}\right)^{3/2}
}\eqno(2.8d).$$
Here the fermion Yukawa coupling is $\rm \lambda_{f} =
m_{f}/v$ where $v = 250$GeV and $\rm m_{f}$ is the fermion
mass. $\rm m_{h}$ is the Higgs boson mass, and $\rm
\Gamma_{h}$ is the Higgs decay width, for which we use the
standard model values$\rm ^{[19]}$.
We should note that the assumption made in arriving at
(2.7), that $\rm <\sigma_{ann} v_{rel}>$ is T independent,
is strictly true only for freeze-out temperatures small
compared with the electroweak phase transition temperature
$\rm T_{EW}$, where$\rm ^{[20]}$ 
$${\rm T_{EW} = \frac{2.4
m_{h}}{\left(1+0.62\left(\frac{m_{t}}{m_{W}}\right)^{2}
\right)^{1/2}}  }\eqno(2.9),$$
and $\rm m_{t}$ is the t quark mass.
The thermal expectation value of the Higgs field is given by
$\rm <h^{o}>_{T} =
v\left(1-\frac{T^{2}}{T_{EW}^{2}}\right)^{1/2}$.  
Thus for  
$\rm T_{fS} \; ^{>}_{\sim}\; T_{EW}$ the effective mass of
the W and Z bosons goes to zero, while the $\rm <h^{o}>_{T}$
dependent S and $\rm h^{o}$ masses differ from their zero
temperature values. In practice, however, $\rm T_{fS} \;
^{>}_{\sim}\; T_{EW}$ occurs only for very large S masses
$\rm ^{>}_{\sim} \; 1TeV$. In this limit, the S mass is 
essentially determined by the constant mass term m in (2.1)
and so is effectively $\rm <h^{o}>$ independent, while from
(2.8b) and (2.8c), in the limit $\rm m_{S} \gg m_{W}$ and
$\rm m_{S} \gg m_{h}/2$ the contribution from 
S annihilations to W and Z bosons reduces to 3 times the
contribution from annihilations to the Higgs boson (2.8a),
as expected in the SU(2)xU(1) symmetric limit. This is T
independent. A large Higgs boson mass doesn't alter this
conclusion, since from (2.9) $\rm T_{EW}$ is of order the
Higgs mass, and so if $\rm T_{fS}$ is of order $\rm T_{EW}$,
then $\rm m_{S}$, which, as discussed below, is much larger than
$\rm T_{fS}$, will also be much larger than $\rm m_{h}$. As
a result, we can neglect $\rm m_{h}$ in the propagators of
(2.8). Thus in practice we can use the cross-sections (2.8)
calculated with the T=0 value for $\rm <h^{o}>$, $\rm
<h^{o}> = 250GeV$.

         Using these contributions to $\rm <\sigma_{ann}
v_{rel}>$ we solve (2.6) self-consistently for the
freeze-out temperature and then obtain from (2.7) the
resulting dark matter density. 
In Figures 2(a)-2(d) we give plots of the dark
matter density as a function of $\rm m_{S}$ and $\rm
\lambda_{S}$ for various values of $\rm m_{h}$. In these we
show the contours for $\rm \Omega_{S} h^{2} = 1.0$,
corresponding to the upper limit from the age of the
Universe, $\rm \Omega_{S} h^{2} = 0.25$, which is the
smallest value for which a critical density ($\rm \Omega_{S} 
= 1$) of S scalars can occur, and $\rm \Omega_{S} h^{2} =
0.025$, corresponding to the smallest value for which S dark 
matter could make up the primary component of the galactic
halo.  In Table 1 we give values of the freeze-out
temperature for various values of
$\rm \lambda_{S}$, $\rm m_{S}$ and $\rm m_{h}$.
Typically $\rm m_{S} = (10-30)T_{fS}$ for the range of
parameters we are considering. We have assumed $\rm m_{t}
=120GeV$ throughout. We find that increasing the t quark
mass to 200GeV makes only a very small difference to the
results.

\vspace{0.2cm}

\bf Table 1. S freeze-out temperature, $\rm
x_{fs}^{-1} = m_{S}/T_{fS}$ }
\newline $\rm \begin{array}{llllllll}
\lambda_{S} & m_{S} & m_{h} & x_{fS}^{-1} & \lambda_{S} &
m_{S} & m_{h} & x_{fS}^{-1} \\
1 & 30 & 60 & 34.5 & 0.01 & 30 & 60 & 25.5\\
1 & 30 & 100 & 23.5 & 0.01 & 30 & 100 & 14.6\\
1 & 30 & 300 & 18.5 & 0.01 & 30 & 300 & 9.6\\
1 & 30 & 500 & 16.5 & 0.01 & 30 & 500 & 7.7\\
1 & 100 & 60 & 27.9 & 0.01 & 100 & 60 & 18.9\\
1 & 100 & 100 & 28.3 & 0.01 & 100 & 100 & 19.3\\
1 & 100 & 300 & 26.5 & 0.01 & 100 & 300 & 17.5\\
1 & 100 & 500 & 23.9 & 0.01 & 100 & 500 & 14.9\\
1 & 1000 & 60 & 25.1 & 0.01 & 1000 & 60 & 16.1\\
1 & 1000 & 100 & 25.1 & 0.01 & 1000 & 100 & 16.1\\
1 & 1000 & 300 & 25.1 & 0.01 & 1000 & 300 & 16.1\\
1 & 1000 & 500 & 25.2 & 0.01 & 1000 & 500 & 16.2\\
\end{array}$
      
\vspace{0.2cm}

       From Figures 2(a)-2(d) we see that for large values of
$\rm \lambda_{S}$ (larger than 0.1), which is particularly
interesting from the point of view of the phenomenology of S scalars, 
in order to have a density of S scalars which can account for a critical 
density of dark 
matter ($\rm \Omega_{S}h^{2} \gae 0.25$), 
we require the S mass typically to be $\rm
^{<}_{\sim}$ 100GeV  
or $\rm ^{>}_{\sim} \; 500GeV$. More generally, as the Higgs
boson mass increases, the value of $\rm \lambda_{S}$ for which S
particles of mass less than about 100GeV can account for
dark matter increases, from about $\rm \lambda_{S} =
0.01-0.1$ for $\rm m_{h} = 
60GeV$ to $\rm \lambda_{S} \; ^{>}_{\sim} \;1$ for $\rm
m_{h} \; ^{>}_{\sim} \; 300GeV$. This could be important
with respect to the possibility of producing S particles via
Higgs decay at future multi-TeV hadron colliders such as the 
CERN Large Hadron Collider (LHC) $\rm ^{[19]}$.  

                  These results are for the case N=1. If we
consider several scalars of equal mass and coupling strength 
to the Higgs (for example, if the $\rm S_{i}$ were a
multiplet under a global symmetry or even gauge symmetry
with a symmetry breaking scale large compared with $\rm
m_{W}$), then it is easy to see that the total density in
$\rm S_{i}$ and $\rm S_{i}^{\dagger}$ is just the sum over
each individual $\rm S_{i}$ density, since each $\rm S_{i}$
annihilates only with its own antiparticle. Thus
$${\rm \Omega_{S\;Total} \equiv \Sigma_{i} \Omega_{S_{i}}  
= N \Omega_{S}   }\eqno(2.10)$$
Since $\rm \Omega_{S} \propto \frac{1}{<\sigma_{ann}
v_{rel}>}
\propto \frac{1}{\lambda_{S}^{2}} $, we see that Figure 2
still holds if we replace $\rm \lambda_{S}$ by $\rm
\hat{\lambda}_{S} = \lambda_{S}/\sqrt{N}$ on the horizontal
axis. Thus for a given value of $\rm \Omega_{S}$ and $\rm
m_{S}$ the value of $\rm \lambda_{S}$ is increased by $\rm
\sqrt{N}$. This will increase strength of interaction with
matter and so the observability of S dark matter for $\rm N
> 1$.

\section{Elastic scattering of S dark matter particles from
nuclei and constraints from Ge detectors.}

          In this section we consider the constraints on
$\rm \lambda_{S}$ and $\rm m_{S}$ following from direct
detection of S dark matter particles via elastic scattering of S
scalars from Ge nuclei$\rm ^{[10,11]}$. It will be assumed
throughout that S dark matter accounts for the halo dark
matter density. Although the simplest possibility for the origin
of a relic density of S particles is from S freeze-out,
in principle there are other possibilities. For example, if
a heavy particle such as a heavy right-handed neutrino N
decays to S particles (via the Higgs-mediated process
$\rm N \rightarrow \nu_{L}S^{\dagger}S$ 
in the case of right-handed neutrinos) at a temperature below the S
freeze-out temperature (typically between 1GeV and 50GeV
for $\rm 20GeV \; ^{<}_{\sim} \; m_{S} \; ^{<}_{\sim} \;
1000GeV$) then the S particles so produced will not return to
an equilibrium density and will result in a relic S density
different from the thermal relic density. In this case  
halo dark matter could, in principle, be accounted for by 
any combination of $\rm m_{S}$ and $\rm \lambda_{S}$. Thus
it is important to consider generally what constraints on
the parameters of the model are imposed by experimental
observations, as well as to compare the constraints with the
thermal relic density as a particular example.

          The S scattering cross-section from quarks via
Higgs exchange gives an effective interaction
$${\rm L_{eff} = \frac{\lambda_{S} m_{q}}{m_{h}^{2}}
\; S^{\dagger}S\overline{q}q    }\eqno(3.1).$$
Using the expression for the nuclear matrix element$\rm^{[16,21]}$
$$\rm <N|\Sigma_{q} \;m_{q}\overline{q}q|N> =  
(7.62) \frac{2}{27}m_{N}\overline{\psi}_{N}\psi_{N}  ~,$$ 
we see that the effective interaction with a
nucleus is given by
$${\rm   L_{eff} = (7.62) \frac{2}{27} \frac{\lambda_{S}
m_{N}}{m_{h}^{2}} S^{\dagger}S\overline{\psi}_{N}\psi_{N} 
        }\eqno(3.2)$$
and that the cross-section for coherent S-nucleus scattering
is given by
$${\rm  \sigma_{S-N} =
(7.62)^{2} \frac{1}{\left(27\pi\right)^{2}} \frac{\pi
m_{N}^{4}}{ \left(m_{S}+m_{N}\right)^{2} }
\frac{\lambda_{S}^{2}}{m_{h}^{4}} }\eqno(3.3).$$

    In general $\rm \sigma_{S-N}$ must be multiplied by a
correction factor $\rm \zeta_{N}(m_{S})$, which accounts for 
the fact that at large enough momentum transfer the
scattering ceases to be a coherent scattering with the whole 
nucleus$\rm ^{[14]}$. We will use a correction factor based
on integrating a Gaussian nuclear form factor over the
Maxwellian velocity distribution of the halo dark matter
particles $\rm ^{[14,22]}$:
$${\rm  \zeta_{N}(m_{S}) = \frac{0.573}{b} \left[1 -
\frac{exp(-\frac{b}{(1+b)})}{\sqrt{(1+b)}}
\frac{erf(\sqrt{\frac{1}{1+b}})}{erf(1)} \right]
}\eqno(3.4),$$ 
where 
$${\rm b =  
\frac{8}{9} \overline{v}^{2} r_{charge}^{2}
\frac{m_{S}^{2}m_{N}^{2}}{(m_{S}+m_{N})^{2}}
}~,\eqno(3.5)$$
$$\rm r_{charge} = 5.1 (0.3+0.89 A^{1/3})GeV^{-1} ~,$$ 
and $\rm \overline{v} = v_{300} 300kms^{-1}$ is the mean halo
velocity dispersion of the S particles, which is related to the
galactic rotation velocity in the isothermal sphere model, $v_{rot}$,
by $\overline{v} = \sqrt{3/2} v_{rot}$.  
For the case of scattering from Ge we find that the full
cross-section is given by
$${\rm \sigma_{S-Ge} =  5.7x10^{-36}cm^{-2}
\frac{\zeta_{N}(m_{S})
}{\left(1+\frac{m_{S}}{76GeV}\right)^{2}}
\left(\frac{100GeV}{m_{h}}\right)^{4}
\left(\frac{\lambda_{S}}{0.1}\right)^{2}    }\eqno(3.6),$$
where $\rm \zeta_{N}(m_{S}) $ is given by (3.4) with 
$$\rm b = 2.2 v_{300}^2/(1+76GeV/m_{S})^{2} ~.$$
 In order to
compare with experiments we need the rate of interaction of
the halo dark matter particles with a detector per kg per
day. This is given by (without energy threshold)$\rm
^{[23]}$ 
$${\rm R = \left(\frac{8}{3\pi}\right)^{1/2} 
\;
\frac{\eta_{v} \overline{v}\rho_{h}\sigma_{S-N}\eta_{N}(m_{S})}{m_{S}
m_{N}} = 0.069 \;\rho_{0.4}\; v_{300}\; \sigma^{N}_{36}\;
\left(\frac{100GeV}{m_{N}}\right)
\left(\frac{100GeV}{m_{S}}\right) kg^{-1}d^{-1}
}\eqno(3.7),$$
where $\rm \rho_{0.4}$ is the density of S scalars in the
halo ($\rm \rho_{h}$) in units of 0.4GeV $\rm cm^{-3}$, 
$\rm \eta_{v} \approx 1.3$ is a
correction for the motion of the Sun and the Earth, and $\rm
\sigma^{N}_{36}$ is the corrected S-N cross-section in units 
of $\rm 10^{-36}cm^{2}$. 

    It should be noted that the correction factor $\rm
\zeta_{N}(m_{S})$ is not accurate for dark matter particle
masses much larger than 100GeV $\rm ^{[22]}$. However, we
will see that the experimental constraints in the present
model are most important for S masses less than about
100GeV, in which case the correction factor (3.4) is
accurate to about 10$\rm \%$ $\rm ^{[22]}$. 

      In Figure 3 we show the event rate as a function of
$\rm m_{S}$ and $\rm \lambda_{S}$ for the cases $\rm m_{h} =
60GeV$, 100GeV and 300GeV. We also show the contours
corresponding to the thermal relic S dark matter region of
the parameter space, $\rm 0.025 \; ^{<}_{\sim} \; \Omega_{S} 
\;
^{<}_{\sim} \; 1$. (We have assumed $\rm \rho_{0.4} =
v_{300} = 1$ throughout).   

      The present experimental upper bound on R corresponds
approximately to 100$\rm kg^{-1}d^{-1}$ for $\rm m_{S} \;
^{>}_{\sim} \; 10GeV$ $\rm ^{[10,11,24]}$. In general, present 
ionization detectors may be able to achieve a sensitivity of 
about 10$\rm kg^{-1}d^{-1}$ $\rm ^{[24]}$, while in the
future cryogenic Ge detectors (such as a proposed $\rm 500g
^{73}Ge + 500g ^{76}Ge$ detector$\rm ^{[25]}$) should be able 
to achieve a sensitivity of 0.1$\rm kg^{-1}d^{-1}$. 
We see from Figure 3 that in order to
constrain the thermal relic S region of parameter space
we need an upper bound on R which is less than $\rm
100kg^{-1}d^{-1}$. For an upper bound on the cross-section
of $\rm 10kg^{-1}d^{-1}$, we can probe a small region of
the thermal relic parameter space corresponding to $\rm
\lambda_{S} \; ^{>}_{\sim} \; 0.06$ and $\rm m_{S} \;
^{<}_{\sim} \; 20GeV$. In order to significantly constrain
the possibility of a critical density of S dark matter, $\rm 
0.25 \; ^{<}_{\sim} \; \Omega_{S}h^{2} \; ^{<}_{\sim} \;
1.0$, we require $\rm R \; ^{<}_{\sim} \; 1kg^{-1}d^{-1}$.
$\rm R \; ^{<}_{\sim} \; 0.1kg^{-1}d^{-1}$ would allow us to 
detect or exclude almost all thermal relic S dark matter for $\rm
m_{S} \; ^{<}_{\sim} \; 50GeV$, while 
$\rm R \; ^{<}_{\sim} \; 0.01kg^{-1}d^{-1}$ would detect
almost all thermal relic S dark matter for $\rm m_{S} \;
^{<}_{\sim} \; 100GeV$. These conclusions for $\rm m_{S}
\; ^{<}_{\sim} \; 100GeV$ are essentially independent of
$\rm m_{h}$, as can be seen by comparing Figures 3(a), 3(b) and
3(c). For $\rm m_{S} \; ^{>}_{\sim} \; 100GeV$, the amount of
thermal relic parameter space which can be
experimentally searched decreases as $\rm m_{h}$ increases.
From Figure 3(a) we see that even with $\rm m_{h}$ as small
as 60GeV, in order to constrain the thermal relic dark
matter for $\rm m_{S} \; ^{>}_{\sim} \; 100GeV$ we would
need $\rm R \; ^{<}_{\sim} \; 0.01kg^{-1}d^{-1}$, and even if an 
upper bound as low as $\rm 0.01kg^{-1}d^{-1}$ could be
achieved, this would not be sufficient to constrain the
possibility of a critical density of thermal relic S dark
matter for $\rm m_{S} \; ^{>}_{\sim} \; 100GeV$.

        Thus we can conclude that since present Ge
ionization detectors give an upper bound on R of about $\rm
100kg^{-1}d^{-1}$ (and are not expected achieve a
sensitivity better that$\rm 10kg^{-1}d^{-1}$), present
attempts at direct detection of dark matter can at best only 
impose a very weak constraint on the possibility of thermal
relic S dark matter. The next generation of cryogenic
detectors should be able to effectively search for S dark
matter up to at least 50GeV. Thermal relic S scalars
significantly heavier than 100GeV are probably beyond the
reach of future Ge detectors, even if the Higgs mass is as
small as 60GeV.

\section{High energy neutrinos from $\rm
SS^{\dagger}$ annihilation in the Earth and the Sun.}

             In this section we calculate the flux of
upward-moving muons and the rate of contained events in
neutrino detectors due to high-energy neutrinos ($\rm >
2GeV$) resulting from annihilations of $\rm SS^{\dagger}$
pairs in the core of the Earth and the Sun$\rm
^{[12-17]}$.  
        
        The rate of upward-moving muons at the surface of
the Earth due to annihilations in the Sun is given by $\rm
^{[13]}$
$${\rm \Gamma_{detector} = 1.27x10^{-29}Cm_{S}^{2}
\Sigma_{i} a_{i}b_{i} \Sigma_{F} B_{F} <N z^{2}>_{F\;i}
m^{-2}yr^{-1}}\eqno(4.1),$$
where C is the capture rate in the Sun in units $\rm
s^{-1}$, $\rm a_{i}$ and $\rm b_{i}$ are the
neutrino-scattering and muon-range coefficients, summed over 
$\rm i = \nu_{\mu}$ and $\rm \overline{\nu}_{\mu}$ ($\rm   
a_{\nu_{\mu}} = 6.8$, $\rm a_{\overline{\nu}_{\mu}} = 3.1$,
$\rm   b_{\nu_{\mu}} = 0.51$, $\rm
b_{\overline{\nu}_{\mu}} = 0.67$) and the $\rm B_{F}$ are
the branching ratios for $\rm SS^{\dagger}$ annihilations to
gauge boson, Higgs boson and quark pairs. $\rm
<Nz^{2}>_{Fi}$ are the second-moments of the spectrum of
neutrino type i from final state F scaled by the S mass
squared,
$${\rm   <Nz^{2}>_{Fi} = \frac{1}{m_{S}^{2}} \int
\left[\frac{dN}{dE}\right]_{Fi} E^{2}dE  }\eqno(4.2),$$
where $\rm \left[\frac{dN}{dE}\right]_{Fi}$ is the
differential energy spectrum of neutrino i at the surface of 
the Sun or Earth resulting from injection of particles in
final state F at the centre of the Sun or Earth. For the
case of annihilations in the Earth one multiplies (4.1) by
$\rm 5.6x10^{8}$, corresponding to the ratio of the distance
squared to the Sun to the radius squared of the Earth $\rm
^{[13]}$.

                 The capture rate is given by$\rm
^{[12,14,16,17]}$ 
$${\rm C = c\frac{\rho_{0.4}}{m_{S} v_{300}}
\Sigma_{N}\sigma_{40}^{N}
F_{N}(m_{S}) f_{N}\phi_{N}S_{N}/m_{N} }\eqno(4.3),$$
where $\rm \sigma_{40}^{N}$ is the S-nucleus elastic
scattering cross-section in units of $\rm 10^{-40}cm^{2}$,
$\rm c = 5.8x10^{24}s^{-1}$ for the Sun and $\rm c =
5.7x10^{15}s^{-1}$ for the Earth. The sum is over all
species of nuclei N in the Earth or Sun. $\rm \phi_{N}$ and
$\rm f_{N}$ are given in Table A.1 of ref.17. $\rm S_{N}$ is a
factor which takes into account the fact that the S dark
matter particle must lose sufficient momentum to be
captured. For $\rm S_{N}$ we have 
$${\rm  S_{N} \approx (1+A_{N}^{-1})^{-1}\;\;\;;\;\;\;
A_{N} = \frac{3}{2}\frac{m_{S}m_{N}}{(m_{S}-m_{N})^{2}}
\left[\frac{v_{esc}^{2}}{\overline{v}^{2}}\right]\phi_{i}
}\eqno(4.4),$$
where $\rm v_{esc}$ is the escape velocity for the Sun
or Earth ($\rm 618kms^{-1}$ and $\rm 11kms^{-1}$
respectively). This has the correct behaviour for $\rm
A_{N}$ large and small compared to 1 $\rm ^{[14,16]}$. $\rm
F_{N}(m_{S})$ is a factor which takes into account form
factor suppression. The branching ratios $\rm B_{F}$,
corresponding to the rates for S annihilation in the limit
of zero relative velocity, are directly obtained from (2.8).

         The rate (4.1) assumes that the accretion of S
particles by the Sun or Earth and their subsequent
annihilation are in equilibrium, in which case the
annihilation rate is given by $\rm \Gamma_{ann} = C/2$. The
condition for this to be true is that the age of the solar
system $\rm t_{\odot}$ be large compared with the time for 
equilibrium to be established $\rm \tau_{A}$, which
is defined below. In general C in (4.1) should be replaced
by$\rm^{[12,16]}$ 
$${\rm C \rightarrow C\;tanh^{2}(\frac{t_{\odot}}{\tau_{A}})
    }\eqno(4.5),$$
where 
$${\rm   \tau_{A} = 1/(CC_{A})^{1/2}\;\;\;;\;\;\;\;
C_{A} = <\sigma v> V_{2}/V_{1}^{2}    }\eqno(4.6).$$
$\rm <\sigma v>$ is the spin-averaged total annihilation
cross-section times the relative velocity in the limit of
zero relative velocity, which can be obtained from (2.8).
The effective volumes $\rm V_{j}$ are given by
 $${\rm V_{j} = 6.5x10^{28}(jm_{S}/10GeV)^{-3/2}\;cm^{3}
}\eqno(4.7a),$$
for the Sun $\rm ^{[12]}$ and 
$${\rm V_{j} = 2.0x10^{25}(jm_{S}/10GeV)^{-3/2}\;cm^{3}
}\eqno(4.7b)$$
for the Earth $\rm ^{[14]}$. 

      A second assumption in obtaining (4.1) is that the capture rate 
is primarily due to single collisions with nuclei ("optically thin" limit). 
However, for the case of capture due to scattering from iron in the Earth, 
it has been pointed out that multiple collisions can enhance the
capture rate $\rm ^{[15]}$. The enhancement factor is given by 
$${\rm \alpha(\tau) = \frac{\exp\left(\tau_{eff} - 1\right)}{\tau_{eff}} \;\;,\;\;\;\; 
\tau_{eff} = \tau \beta_{-}  } ~,\eqno(4.8)$$
where 
$${\rm \tau \approx \sigma_{S-Fe}/\left(2.3 \times 10^{-33} \; cm^{2}\right)   }   $$
is the optical depth of the Earth, 
$\rm \beta_{-} = 4 m_{S}m_{Fe}/\left(m_{S} - m_{Fe}\right)^{2}$, and $\tau_{eff}$ is the effective optical depth of the Earth taking into account multiple collisions. 
This expression is valid so long as $\rm Max(1,ln \beta_{-}) \lae 6/\tau_{eff}$ and 
$\rm \beta_{-} \lae 20$; otherwise the enhancement must be evaluated numerically, although the largest enhancement occurs typically for $\rm \beta_{-} \approx 20$ $\rm ^{[15]}$. We have 
included $\alpha(\tau)$ from (4.8) in our calculations over the range where it is valid (and where enhancement is expected to be most important), in order to indicate the importance or otherwise of multiple collisions. In practice, we find that no enhancement  
of the event rate in detectors occurs over the range of parameters we are considering.

                The $\rm <Nz^{2}>_{Fi}$ are related to the
muon neutrino and anti-neutrino energy spectra coming from
annihilation of S in the Sun and Earth, including in the
case of the Sun the effects of the interactions of the
annihilation products and neutrinos with the solar
medium$\rm ^{[13]}$. For $\rm m_{S} > m_{W}$ the
dominant contributions to the $\rm <Nz^{2}>_{Fi}$ are from
annihilation to gauge boson and t quark final states, while
for $\rm m_{S} < m_{W}$ the dominant final states
contributing to $\rm <Nz^{2}>_{Fi}$ are b quark pairs and
possibly Higgs boson pairs if $\rm m_{h} < m_{W}$. In the
Appendix we discuss the values of 
$\rm <Nz^{2}>_{Fi}$ coming from the different final states.

             In order to calculate the capture rate for the
case of the Earth one can simply use (4.3), since in this
case the form factor suppression is small for most values of $\rm m_{S}$ $\rm ^{[14]}$
and so we can take $\rm F_{N}(m_{S}) = 1$ $\rm ^{[16]}$.
[Capture is dominated by low momentum transfer
scattering excpet for $m_{S}$ close to the mass of the scattering nucleus, in which case the form factor suppression can be more significant (a factor of 0.72 for the case where $m_{S} = m_{Fe}$ $\rm ^{[14]}$).] For the case of capture by the
Sun, form factor suppression cannot be neglected, making the
calculation of the capture rate more complicated. A simple
expression for the capture rate in this case has been given
by Kamionkowski$\rm ^{[16]}$, which is accurate to
$\rm 5\%$ for dark matter particle masses greater than a few 
GeV and less than a few TeV (see also Ref.[14]). In terms of $\rm \sigma_{S-N}$
this may be written as
$${\rm C = \frac{\pi}{4}
\frac{\left(m_{S}+m_{N}\right)^{2}}{m_{S}^{2}m_{N}^{4}}
\; \sigma_{S-N} \; f_{S}(m_{S})    }\eqno(4.9),$$
where
$${\rm \begin{array}{lll} f_{S}(m_{S}) =  & 2.04 \times 10^{38}
\exp[-0.0172(m_{S}-10)], & m_{S} \leq 80 \GeV \\
&
6.10 \times 10^{37}(m_{S}/80)^{-1.06-0.38[(m_{S}-80)/920]^{1/2}},&
80\GeV \leq m_{S} \leq 1000 \GeV\\
& 1.72 \times 10^{36}(m_{S}/1000)^{-1.88}, &  m_{S} \geq
1000 \GeV \\ \end{array}  }\eqno(4.10).$$
Here C is in $\rm s^{-1}$ and all masses are in GeV.

      In Figures 4(a)-4(c) we show the results for S
annihilations in the Sun for the cases $\rm m_{h} = 60GeV$,
$\rm 100GeV$ and $\rm 300GeV$ respectively, and in Figures 
5(a)-5(c) we show the corresponding results for the case of
annihilations in the Earth. Comparing Figures 4 and 5,
 we see that for an upper
bound on $\rm \Gamma_{detector}$ corresponding to the IMB
upper bound$ ^{[26]}$, $\rm \Gamma_{detector} <
2.65x10^{-2}m^{-2} yr^{-1}$ (curve (a) in Figures 4 and 5),
the strongest constraints on the parameter space come from
Earth S annihilations, while for upper bounds less than
around $\rm 10^{-3}m^{2}yr^{-1}$ the solar S annihilations
become more important. Thus we will compare the thermal
relic parameter space with the Earth S annihilation
constraints for 
the case of the present IMB upper bound and with the solar
annihilation S constraints for the case of the bounds
expected from future neutrino detectors. 

           From Figure 5(a), corresponding to Earth S
annihilations with $\rm m_{h} = 60GeV$, we see that at
present the IMB constraints can exclude only a small region
of the thermal relic parameter space (corresponding to the
iron 'resonance' at $\rm m_{S} \approx 56GeV$). From Figure
4(a), we see that for an upper bound on $\rm
\Gamma_{detector}$ of $\rm 10^{-3}m^{-2}yr^{-1}$,
upward-moving muons from solar S annihilations can exclude
a small region of the thermal relic parameter space corresponding to 
$\rm \lambda_{S} \gae 0.1$ and  $\rm m_{S} \;
^{<}_{\sim} \; \; 50GeV$. [We see from Figure 5(a)
that for this case the bounds due to neutrinos from the Earth are stronger that those due to solar neutrinos for $m_{S}$ between 20 and 80 GeV, and can probe a significant region of the thermal relic parameter space for $m_{S}$ between 50 and 70 GeV.] 
With an upper bound $\rm
\Gamma_{detector} <  10^{-4}m^{-2}yr^{-1}$, most of the
thermal relic parameter space in Figure 4(a) corresponding to $\rm
\Omega_{S}h^{2} \; ^{<}_{\sim} \; 0.25$ for $\rm m_{S} \;
^{<}_{\sim} \; 400GeV$ and a significant region of the
thermal relic parameter space corresponding to a critical S
denisty for $\rm m_{S} \; ^{<}_{\sim} \; 50GeV$ can be investigated, 
while an upper bound $\rm \Gamma_{detector} <
10^{-5}m^{-2}yr^{-1}$ would probe the whole thermal relic
parameter space up to $\rm m_{S} \approx 500GeV$. For larger
$\rm m_{h}$, the conclusions for $\rm m_{S} \; ^{<}_{\sim}
\; 100GeV$ are essentially unchanged, while for $\rm m_{S}
\; ^{>}_{\sim} \; 100GeV$ the amount of thermal relic
parameter space which can be investigated for a given upper bound on $\rm
\Gamma_{detector}$ decreases as $\rm m_{h}$ increases. 

            In order to see how the IMB upper bound could be 
improved in the future, we can make a rough estimate of the
bound which could be imposed by building neutrino detectors
of larger area. The IMB bounds follow from a detector area
of $\rm 400m^{2}$ and exposure of about 1 year,
corresponding to an upper bound of less than about 10 upward 
moving muons per year. Following ref.[27], we can estimate the 
area of detector required in order to achieve a given
sensitivity by the area needed to detect one upward-moving
muon event per year. At present, the MACRO detector at Gran
Sasso, with an area $\rm \approx 10^{3}m^{2}$, is beginning
operation$\rm ^{[28]}$, while a number of detectors with an
effective area of order $\rm 10^{4}m^{2}$ are under
development (DUMAND$\rm ^{[29]}$; AMANDA$\rm ^{[30]}$;
NESTOR$\rm ^{[31]}$). In addition, it has been suggested$\rm 
^{[27]}$ that a $\rm 1km^{2}$ detector is needed to observe
muons from neutralino dark matter in the GeV-TeV mass range, 
and that this could be constructed at a cost of order 100
million U.S. dollars $\rm ^{[32]}$. We see from Figure 4
that a detector of area $\rm 10^{4}m^{2}$ should be able to
probe the region of parameter space corresponding to curve 
(c), which will rule out much of the parameter space
corresponding to thermal relic S dark matter with $\rm m_{S} 
\; ^{<}_{\sim} \; 50GeV$. This conclusion is essentially
independent of $\rm m_{h}$, as can be seen by comparing
Figures 4(a), 4(b) and 4(c). For the case 
of a 1$\rm km^{2}$ detector, the area of parameter space
under curve (e) in Figure 4 could in principle be searched.
This would probe the entire thermal relic dark matter
region for $\rm m_{S} \; ^{<}_{\sim} \; 1.5TeV$
(500GeV, 100GeV) for the case of $\rm m_{h} = 60GeV$
(100GeV, 300GeV). Thus in general the thermal relic dark
matter parameter space can be probed for $\rm m_{S}$ at
least up to 100GeV. 

       We therefore conclude that at present the IMB
upper bound on the flux of upward-moving muons can impose
only a slight constraint on the possibility of thermal relic 
S dark matter, while many of the S dark matter possibilities 
with $\rm m_{S} \; ^{<}_{\sim} \; 50GeV$ should be within
the reach of experiments with an area $\rm O(10^{4})m^{2}$ in
the not-too-distant future. In the more distant future large 
($\rm 1km^{2}$) detectors should be able detect or exclude S 
dark matter for $\rm m_{S}$ up to at least 100GeV. 
 
          We also note that comparing the next generation of 
Ge detectors, which might reach scattering rates $\rm
0.1kg^{-1}d^{-1}$ (Figure 3, curve e), with the next
generation of neutrino detectors, which might reach
upward-moving muon rates $\rm 10^{-4}m^{-2}yr^{-1}$       
(Figure 4, curve c), we see that for $\rm m_{S} \;
^{>}_{\sim} \; 80GeV$ the constraint from upward-moving
muons are dominant, while for $\rm m_{S} \; ^{<}_{\sim} \;
80GeV$ Ge detectors impose a stronger constraint.

             Up to now we have only considered the rate of
upward-moving muons in discussing constraints on the $\rm
(\lambda_{S}, m_{S})$ parameter space. The upward-moving
muon flux is the most important signal for high-energy
neutrinos from the point of view of future large-area
neutrino detectors, which are specifically designed to
detect this flux. However, in discussing the present bounds
on the flux of high-energy neutrinos due to $\rm
SS^{\dagger}$ annihilations, we should also consider the
possibility that a high-energy electron or muon neutrino
could undergo a charged current interaction within the
volume of the detector$\rm ^{[13]}$ ('contained event'). For 
the case of neutrinos from the Sun the rate of contained 
events per kiloton due to electron and muon neutrinos is
given by$\rm ^{[13]}$
$${\rm \Gamma_{detector} =
3.3x10^{-27}Cm_{S}\Sigma_{i}a_{i}\Sigma_{F}B_{F}<Nz>_{F\;i}
kt^{-1}yr^{-1}   }\eqno(4.11),$$
where i is summed over the electron and muon neutrino and
anti-neutrino. In the Appendix we discuss the values of $\rm 
<Nz>_{F\;i}$ coming from the various final states. In
Figure 6 we show the results for S annihilations in the
Sun for the cases $\rm m_{h} =$ 60GeV, 100GeV and 300GeV,
while in Figure 7 we show the corresponding results for
the case of S annihilations in the Earth. The present upper
bound on the rate of electron and muon contained events in
the Frejus detector is$\rm ^{[17,33]}$ $\rm 
\Gamma_{detector} < 6.4\;kt^{-1}yr^{-1}$, corresponding to
curve (a) in Figures 6 and 7. Comparing with the upward-moving
muon bounds from Figures 4 and 5, we see that at present the
contained event rate imposes constraints on the parameter
space which are in general weaker than those coming from the 
upward-moving muon flux, although at small $\rm m_{S}$, $\rm
m_{S} \lae 20GeV$, the constraints become similar
(and slightly stronger for the case of solar neutrinos).

        So far in this section and in the previous
section, we have considered the case of just one S scalar.
For the case of N scalars of equal mass and coupling, the
density of each scalar $\rm S_{i}$ contributes a proportion
1/N of the total halo density. The
capture rate of S dark matter in the Sun and the Earth and
the rate of elastic scattering in Ge detectors are
proportional to the density of $\rm S_{i}$ in the halo times 
the cross-section for scattering from nuclei in the Sun and
Earth. Thus the contribution to the event rate in a detector
is reduced by a factor 1/N for a given $\rm S_{i}$. The
total rate from summing over i for a given $\rm m_{S}$ and
$\rm \lambda_{S}$ is therefore unchanged. However, for a
given value of the thermal relic density, the value of $\rm
\lambda_{S}$ for a given $\rm m_{S}$ is increased by a
factor $\rm \sqrt{N}$, leading to an increase in the
scattering cross-section and so to an increase in the event
rate in Ge detectors and in neutrino detectors by a factor N
for a given thermal relic density, thus making the dark
matter easier to detect.

\section{Conclusions}   

               The extension of the standard model by the
addition of a gauge singlet scalar provides a canonically
minimal extension of the standard model which can
potentially account for dark matter. It is important,
therefore, to consider in some detail the question of the
relic density of the gauge singlet scalars and their
possible observable signatures. In general, present
experiments based on observing elastic scattering of 
halo dark matter patricles from Ge nuclei, and on observing
upward-moving muons at the Earth's surface, coming from muon
neutrinos due to dark matter particle annihilation in the
Sun or the Earth, can only place very weak constraints on 
thermal relic S dark matter, and cannot constrain the
possibility that thermal relic S dark matter could account
for a critical density of dark matter ($\rm \Omega_{S} =
1$). However, the next generation of cryogenic Ge detectors
(which hopefully should achieve bounds on the Ge scattering
rate of $\rm 0.1kg^{-1}d^{-1}$) and neutrino detectors (with
an effective area $\rm 10^{4}m^{2}$) will be able to
investigate most of the parameter space for thermal relic S
scalar dark matter with $\rm m_{S} \; ^{<}_{\sim} \; 50GeV$, 
while a 1$\rm km^{2}$ neutrino detector, as suggested in
order to search for heavy neutralino dark matter, would be
able to exclude thermal relic S dark matter for $\rm m_{S}
\; ^{<}_{\sim} \; 100GeV$ (as would a cryogenic Ge detector
if it could achieve a sensitivity of $\rm
0.01kg^{-1}d^{-1}$). For a light Higgs mass, equal to 60GeV
(100GeV), a $\rm 1km^{2}$ detector could also exclude
heavier thermal relic S dark matter up to 1.5TeV (500GeV).
In general, the next generation of cryogenic detectors will be 
the most effective in searching for S dark matter with $m_{S} \lae 80 \GeV$, 
while for $m_{S} \gae 80 \GeV$ the next generation of neutrino detectors will 
be most effective.   

           The coupling of a gauge singlet scalar to
the standard model Higgs doublet is unique in form and
inevitably will be a feature of many particle physics models
beyond the standard model. We believe the results presented
here may generally be useful in the study of such models and 
of their cosmological consequences.

\section*{Appendix. $\rm
<Nz^{2}>_{F\;i}$ and $\rm <Nz>_{F\;i}$ from S$\rm
{\bf S^{\dagger}}$ annihilations.} 

           In this Appendix we give the dominant
contributions to $\rm <Nz^{2}>_{Fi}$ and $\rm
<~Nz~>_{Fi}$ for the gauge boson, Higgs boson and quark
final states coming from $\rm SS^{\dagger}$ annihilation. We will use the discussion of Ritz and Seckel $\rm ^{[13]}$ (RS) for the case of the quark final states, while for the case of the gauge boson final states we will follow Ref.[16] and consider $\rm <Nz^{2}>_{Fi}$ and $\rm <~Nz~>_{Fi}$ to mostly originate from the highest-energy "semiprompt" W and Z decays to neutrinos. For the case of the Higgs boson final states we will adapt the results of RS to obtain $\rm <Nz^{2}>_{Fi}$ and $\rm <~Nz~>_{Fi}$.  

\begin{center} $\rm {\bf SS^{\dagger} \rightarrow WW,ZZ}$ \end{center}

           In this case the dominant contribution to
$\rm <Nz^{2}>_{F \; \nu_{\mu}} $ comes from muon neutrinos
originating in the decays $\rm W^{+} \rightarrow \mu^{+}
\nu_{\mu}$ and 
$\rm Z^{o} \rightarrow \nu_{\mu} \overline{\nu}_{\mu}$. The
mean energy squared of the neutrinos is $\rm E_{o} = 
\frac{m_{S}^{2}}{4} (1+\beta^{2}/3)$, where $\rm \beta$ is
the velocity of the decaying W or Z$\rm ^{[16]}$ [$\rm \beta
= (1-m_{W}^{2}/m_{S}^{2})^{1/2}$ for the case of the W].
This assumes that in the rest frame of the W, the W decays
isotropically to final states each of energy $\rm m_{W}/2$.
The branching ratio of $\rm W^{+}$ to $\rm \nu_{\mu}$ decays 
is given by
1 divided by the number of SU(2) doublets to which W can
decay, which gives 1/9 for W decaying to all lepton doublets
and 1st and 2nd generation quark doublets. Thus, noting that N is the number of neutrinos produced per injected boson or fermion pair $\rm ^{[13]}$, we see that the 
$\rm <Nz^{2}>_{Fi}$ following from annihilation to W
pairs can be estimated to be
 $${\rm <Nz^{2}>_{W \; \nu_{\mu}} \approx \frac{1}{9}\;
\frac{1}{4}
(1+\beta^{2}/3) =  0.028(1+\beta^{2}/3)     }\eqno(A.1).$$
For the case of annihilation to Z pairs, the branching ratio for $\rm Z \rightarrow 
\overline{\nu}_{\mu} \nu_{\mu}$ is 0.066 $\rm ^{[19]}$, and so the $\rm <Nz^{2}>_{Fi}$
can be estimated to be
 $${\rm <Nz^{2}>_{Z \; \nu_{\mu}} \approx 2(0.066)\frac{1}{4}
(1+\beta^{2}/3) = 0.033(1+\beta^{2}/3)      }\eqno(A.2),$$  
where the factor 2 occurs because
either of the Z's produced by S annihilation can lead to a 
$\rm \nu_{\mu}$. The same results are obtained for $\rm i=
\overline{\nu}_{\mu}$. For the case of the $\rm <Nz>_{Fi}$
one obtains in the same way, for $\rm i = $ e and $\rm \mu$, 
 $${\rm <Nz>_{W \; \nu_{i}} \approx \frac{1}{9} \frac{1}{2} = 0.056
}\eqno(A.3)$$
and
$${\rm <Nz>_{Z \; \nu_{i}} \approx 2(0.066)\frac{1}{2} = 0.066
}\eqno(A.4),$$
where we have replaced the mean squared energy of the neutrinos  
$\frac{m_{S}^{2}}{4} (1+\beta^{2}/3)$ in (A.1) and (A.2) by the mean squared energy in the rest frame $\rm m_{S}/2$. The values of $\rm <Nz>_{Fi}$
for $\rm \overline{\nu}_{i}$ are equal to those of $\rm \nu_{i}$.

These results are true for the case
where interactions of the W, Z and neutrinos with the Sun or Earth
are ignored. This is justified for the Earth, but for the
case of the Sun there is an additional suppression factor
due to the absorption of muon neutrinos (due to charged
current interactions) and loss of neutrino energy (due to
neutral current interactions) as the neutrinos pass through
the Sun $\rm ^{[13]}$. (The W and Z will decay fast enough that the effect of their interaction with the solar medium prior to their decay can be ignored $\rm ^{[13]}$.)
The suppression factors are given
by$\rm ^{[13]}$ 
$${\rm P_{i} = 1/(1+E_{o}\tau_{i})^{n+\alpha_{i}}   
}\eqno(A.5),$$
where $\rm E_{o}$ is the initial neutrino energy and $\rm n = $ 2(1) 
for the case of  $\rm <Nz^{2}>_{Fi}$ ($\rm <Nz>_{Fi}$). $\rm
\alpha_{\nu_{\mu}} = 5.1$, $\rm
\alpha_{\overline{\nu}_{\mu}} = 9.0$, $\rm \tau_{\nu_{\mu}}
= 1.01x10^{-3}GeV^{-1}$ and $\rm \tau_{\overline{\nu}_{\mu}} 
= 3.8x10^{-4}GeV^{-1}$ for $\rm i =$ e, $\rm \mu$. 
The unsuppressed $\rm <Nz^{2}>_{Fi}$ and  $\rm <Nz>_{Fi}$
are multiplied by
the $\rm P_{i}$ in order to obtain the true $\rm
<Nz^{2}>_{Fi}$ and  $\rm <Nz>_{Fi}$
for the case of neutrinos from the Sun.  

          It is important to note that the assumption 
that the  $\rm <Nz^{2}>_{Fi}$ and $\rm <Nz>_{Fi}$  
are dominated by the "semiprompt" decays of the W and Z is well justified for the case of the unsupressed  $\rm <Nz^{2}>_{Fi}$ and $\rm <Nz>_{Fi}$  $\rm ^{[16]}$, which is appropriate for the case of neutrinos from the Earth. However, for the case of neutrinos 
from the Sun, because the the higher-energy neutrinos from semiprompt decays are preferentially absorbed relative to the lower-energy neutrinos 
coming from secondary decays $\rm ^{[13]}$ (such as W's decaying to pairs of quarks which subsequently decay to neutrinos), the secondary decay neutrinos may become important at large S masses. At the end of this Appendix we make an estimate of the importance of the secondary decays for the case of the Z boson final state, where it is shown that the primary decays dominate  $\rm <Nz^{2}>_{F\nu}$ ($\rm <Nz^{2}>_{F\overline{\nu}}$) for
$\rm m_{S}$ up to at least 1.4 TeV (2.2 TeV), and up to at least 860 GeV (1.3 TeV) for  $\rm <Nz>_{F\nu}$ ($\rm <Nz>_{F\overline{\nu}}$). From the Figures we see that an underestimate of the  $\rm <Nz^{2}>_{Fi}$ or $\rm <Nz>_{Fi}$ by a factor of 2 will make very little difference to our conclusions. Thus we expect that in general our results for the case of solar $S$ annihilations will be reliable for $\rm m_{S}$  up to at least $\sim $ 1.5 TeV for the upward-moving muons and up to at least $\sim$ 1 TeV for the contained events. 

\begin{center} $\rm {\bf SS^{\dagger} \rightarrow
\overline{t}t\;\;,\;\;\overline{b}b}$ \end{center}

          In the case of quark final states, one must consider the details of hadronization and fragmentation of the final state quarks, which will produce hadron jets. 
RS $\rm ^{[13]}$ have used the results of the Lund Monte Carlo program, which simulates the final states of $\rm e^{+}$, $\rm e^{-}$ annihilations into fermion pairs, in order to calculate the values of $\rm <Nz^{2}>_{F\;i}$  and  $\rm <Nz>_{F\;i}$ due to dark matter particles annihilating to fermion pairs. For the case of non-interacting final state quarks (appropriate for S annihilations in the Earth), one can use the RS results directly. In general, the $\rm <Nz^{2}>_{F\;i}$ and $\rm <Nz>_{F\;i}$ are given by $\rm ^{[13]}$ 
$${\rm <Nz^{2}>_{F\;i}  =  \frac{N}{3} < y^{2} > (<z_{F}^{2}> - \frac{1}{4} z_{M}^{2} ) } 
\eqno(A.6)$$
and 
$${\rm <Nz>_{F\;i}  =  \frac{N}{2} <y><z_{F}> } 
\eqno(A.7),$$
where $\rm z_{M}^{2} = m_{H}^{2}/m_{S}^{2}$, N, $\rm <y^{n}>$, the hadron mass
$\rm m_{H}$, and $\rm <z_{F}^{n}>$ are given in Tables 2 and 3 of Ref. [13]. 
In this we have assumed that the mean hadron energy (scaled by the S mass) whrn the hadron decays, $\rm <z_{H}>$, is equal to the hadron energy after fragmentation $\rm <z_{F}>$, which is true if the hadrons are not slowed by the astrophysical medium (Sun or Earth) before they decay. We will show below that this is in general justified for the case of interest to us here. One has to correct (A6) and (A7) for the case of $m_{S}$ near 
the thershold for producing a hadron, since in this case energy conservation implies 
$\rm <z_{F}> \rightarrow 1$. RS make the replacement $\rm <z_{F}^{n}> \rightarrow 
<z_{F}^{n}> + (1 - <z_{F}^{n}>)z_{M}^{n}$ in order to take this into account $\rm ^{[13]}$. Using (A6) and (A7) 
(corrected for thresholds) and the results of Ref. [13] we obtain, for the $\rm \overline{t}t$ final state, 
$${\rm <Nz^{2}>_{t \nu_{\mu}}  =  1.7 \times 10^{-2} (1 - 0.04 z_{M}^{2} ) } 
\eqno(A.8)$$
and 
$${\rm <Nz>_{t \nu_{i}}  =  4.7 \times 10^{-2} (1 + 0.41 z_{M}^{2}) } 
\eqno(A.9),$$
where $\rm i = e$ or $\mu$. The same results are obtained for the antineutrinos. For 
the $\rm \overline{b}b$ final state we obtain 
$${\rm <Nz^{2}>_{b \nu_{\mu}}  =  6.5 \times 10^{-3} (1 + 0.39 z_{M}^{2} ) } 
\eqno(A.10)$$
and 
$${\rm <Nz>_{b \nu_{i}}  =  2.8 \times 10^{-2} (1 + 0.41 z_{M}^{2}) } 
\eqno(A.11).$$
These results are for the case where the interactions with the astrophysical medium are ignored. For the case of solar annihilations one has to consider the possible effects of hadrons slowing before they decay, as well as the effect of neutrinos losing energy or being absorbed as they pass through the Sun. In fact, we can ignore the effect of hadrons slowing for the case of interest to us here. For the b quark final state, the effect of slowing is only important for $\rm E_{b} > E_{b}^{c} = 470 GeV$ $\rm ^{[13]}$. But the b quark final state is important only when the W and Z final states are kinematically disallowed, $\rm m_{S} < m_{W}$, in which case we can ignore the slowing of the hadrons. 
For the case of the t quarks, one has $\rm E_{t}^{c} = (m_{t}/m_{b})^{1/2}E_{b}^{c} = 2.3 TeV$ for $\rm m_{t} = 120 GeV$. For $\rm m_{S}$ large compared with $\rm m_{W}$, the branching ration to the W final state is much larger than that to the t quark final state. 
[From (2.8) we find $\rm B_{WW}/B_{\overline{t}t} \approx 2 m_{S}^{2}/3m_{t}^{2}$ in the limit of large $\rm m_{S}$.] Thus we see that for values of $\rm m_{S}$ for which the slowing of t quarks becomes important (greater than 1 TeV), we can ignore the t quark final states. Therefore in general we can ignore the effect of quarks slowing before they decay. 

         In order to take account of the interaction of the neutrinos with the Sun, we use the method of RS. We simply integrate the differential energy spectrum, including the $\rm P_{i}$ factors from (A.5): 
$${\rm   <Nz^{n}>_{Fi\;A} = \int \left[\frac{dN}{dz}\right]_{Fi} \frac{z^{n}dz}{\left(1 + 
z/z_{Si}\right)^{n + \alpha_{i}} }    } \eqno(A.12),$$
where  $\rm z_{Si} = 1/\tau_{i}m_{S}$ and $\rm <Nz^{n}>_{Fi\;A}$ is the moment of the neutrino distribution including the effect of interactions with the Sun. For $(n + \alpha_{i})z/z_{Si}$ small compared with 1, the denominator can be expanded to give 
$${\rm    <Nz^{n}>_{Fi\;A} =  <Nz^{n}>_{Fi}\left[ 1 - \frac{<z^{n+1}>}{<z^{n}>} \frac{n + \alpha_{i}}{z_{Si}} \right] }  \eqno(A.13).$$  
Using Table 3 and Eq.(32) of Ref.[13] we find that $\rm (<z>, <z^{2}>,<z^{3}>)$ 
equals $\rm (0.13,4.4 \times 10^{-2},2.1 \times 10^{-2})$ for the t quark final state and 
$\rm (0.13,2.9 \times 10^{-2},9.5 \times 10^{-3})$ for the b quark final state. Thus we 
obtain, for the case of the t quark final state, 
$${ \rm <Nz^{2}>_{t \nu_{\mu}\;A} = <Nz^{2}>_{t \nu_{\mu}} \left[1 - m_{S}/(290\;GeV) \right]    } \eqno(A.14),$$ 
$${ \rm <Nz^{2}>_{t \overline{\nu}_{\mu}\;A} = <Nz^{2}>_{t \overline{\nu}_{\mu}} \left[1 - m_{S}/(492\;GeV) \right]    } \eqno(A.15),$$ 
$${ \rm <Nz>_{t \nu_{\mu}\;A} = <Nz>_{t \nu_{\mu}} \left[1 - m_{S}/(478\;GeV) \right]    } \eqno(A.16),$$ 
$${ \rm <Nz>_{t \overline{\nu}_{\mu}\;A} = <Nz>_{t \overline{\nu}_{\mu}} \left[1 - m_{S}/(765\;GeV) \right]    } \eqno(A.17),$$ 
and, for the b final state, 
$${ \rm <Nz^{2}>_{b \nu_{\mu}\;A} = <Nz^{2}>_{b \nu_{\mu}} \left[1 - m_{S}/(422\;GeV) \right]    } \eqno(A.18),$$ 
$${ \rm <Nz^{2}>_{b \overline{\nu}_{\mu}\;A} = <Nz^{2}>_{b \overline{\nu}_{\mu}} \left[1 - m_{S}/(716\;GeV) \right]    } \eqno(A.19),$$ 
$${ \rm <Nz>_{b \nu_{\mu}\;A} = <Nz>_{b \nu_{\mu}} \left[1 - m_{S}/(740\;GeV) \right]    } \eqno(A.20),$$ 
$${ \rm <Nz>_{b \overline{\nu}_{\mu}\;A} = <Nz>_{b \overline{\nu}_{\mu}} \left[1 - m_{S}/(1200\;GeV) \right]    } \eqno(A.21).$$ 
These should be accurate so long as the suppression factors are not too small compared with 1. However, for the case of the t quark we see that for $\rm <Nz^{2}>_{t \nu_{\mu}\;A} $ the approximation breaks down for $\rm m_{S}$ larger than about 250 GeV. In this case an alternative method for estimating the suppression of neutrinos must be used. From Table 3 of Ref.[13] we see that the effect of fragmentation of the t quark is quite small, 
with $\rm <z_{F}> = 0.87$ and  $\rm <z_{F}^{2}> = 0.78$, compared with 1 for the case without fragmentation. In addition, most of the neutrinos come from the primary decay mode to neutrinos, $\rm t \rightarrow b \mu^{+} \nu_{\mu}$ $\rm ^{[13]}$. This can be seen by comparing the naive estimate based on this decay mode with the results of (A.8) and (A.9). 
Assuming that in the rest frame of the decaying quark the decay is isotropic with each decay product having energy $\rm \approx m_{t}/3$, the energy squared of the neutrino is 
$\rm (m_{S}^{2}/9)(1 + \beta^{2}/3)$, where $\rm \beta = (1 - m_{t}^{2}/m_{S}^{2})^{1/2}$. 
The branching ration for this decay is 1/9. Thus we obtain
$${\rm <Nz^{2}>_{t \nu_{\mu}}  \approx   \left(\frac{1}{9}\right)^{2}
 (1 + \beta^{2}/3) = 0.012 (1 + \beta^{2}/3)  } 
\eqno(A.22),$$
and 
$${\rm <Nz>_{t \nu_{\mu}}  \approx   \frac{1}{9} \frac{1}{3} = 0.037  } 
\eqno(A.23),$$
which in the limit $\rm \beta \rightarrow 1$ are close to (A.8) and (A.9). 

   Thus in this case a reasonable approximation to the suppression factors is to use the 
$\rm P_{i}$ with $\rm E_{o} = m_{t}/3$. At large values of $\rm m_{S}$, where such an approach may fail (due to the preferential stopping of the higher-energy primary decay neutrinos, such that the spectrum is not dominated by these neutrinos $\rm ^{[13]}$), the t quark final state can be neglected compared with the gauge boson final state when calculating event rates. Thus we will use (A.18) - (A.21) for the 
b quark final state and the $\rm P_{i}$ suppression factors for the t quark final state.

\begin{center} $\rm {\bf SS^{\dagger} \rightarrow h^{o}h^{o}} $ \end{center} 

  RS do not explicitly discuss this case. However, we can easily adapt their results. The main decay mode of the Higgs boson when $\rm m_{h} < m_{W}$ (with branching ratio $\approx 0.9$) is to $\rm \overline{b}b $ pairs. (The Higgs boson final state can in general be neglected compared with the gauge boson final states when these are kinematically allowed.) The neutrinos occur in the decay of these $\rm \overline{b}b$ pairs. We can simply regard the decay of the $\rm h^{o}h^{o}$ pair as the injection of two $\rm \overline{b}b$ pairs, with each b quark having a mean energy $\rm m_{S}/2$. This should be a good approximation for $\rm m_{S}/2 \gg m_{b}$ In this case we can use the RS results for $\rm \overline{b}b$ pairs, but with $\rm m_{S} \rightarrow m_{S}/2$ and an overall factor of 2. This gives, for the non-interacting case, 
$${\rm <Nz^{2}>_{h^{o} \nu_{\mu}}  =  3.3 \times 10^{-3} (1 + 1.6 z_{M}^{2} ) } 
\eqno(A.24)$$
and 
$${\rm <Nz>_{h^{o} \nu_{i}}  =  2.8 \times 10^{-2} (1 + 0.82 z_{M}^{2}) } 
\eqno(A.11).$$
The suppression factors for the interacting case are $\rm [1 -m_{S}/(844 GeV)] $ 
($\rm \nu$) and $\rm [1 -m_{S}/(1.4 TeV)] $ 
($\rm \overline{\nu}$) for the $\rm <Nz^{2}>_{Fi}$ and     
$\rm [1 -m_{S}/(1.5 TeV)] $ 
($\rm \nu$) and $\rm [1 -m_{S}/(2.4 TeV)] $ 
($\rm \overline{\nu}$) for the $\rm <Nz>_{Fi}$. 

       We can also use this method to estimate the contribution of the secondary decay neutrinos to the $\rm <Nz^{2}>_{Fi}$ and $\rm <Nz>_{Fi}$ for the case of solar S annihilations to gauge boson final states. For the case of S annihilations to a pair of Z bosons, the secondary neutrinos come from the decay of the Z's to a $\rm \overline{b}b$, 
$\rm \overline{c}c$, or  $\rm \overline{\tau}\tau$ pair. (Other lighter quarks or leptons are stopped in the Sun prior to their decay and can be neglected $\rm ^{[13]}$.) Thus we can use the results of RS for the case of injection of a pair of b or c quarks or $\rm \tau$ leptons each of energy $\rm m_{S}/2$. (For the b or c quarks, this will overestimate the contribution when $\rm m_{S}/2 > E_{b}^{c}$ or $\rm E_{c}^{c}$, since we are then neglecting the slowing of the b and c quarks prior to their decay.) The branching ration for Z decay is 0.15 to a b quark pair, 0.12 to a c quark pair, and 0.033 to a $\rm \tau$ lepton pair $\rm ^{[19]}$. Thus we find, using the results of RS, that the contribution of the secondary decays is given by 
$${\rm <Nz^{2}>_{Zb \nu}  \approx  4.9 \times 10^{-4}   }\eqno(A.26a),$$   
$${\rm <Nz^{2}>_{Zc \nu}  \approx  1.7 \times 10^{-4}   }\eqno(A.26b),$$   
$${\rm <Nz^{2}>_{Z\tau \nu}  \approx  3.8 \times 10^{-4}   }\eqno(A.26c),$$   
and
$${\rm <Nz>_{Zb \nu}  \approx  4.2 \times 10^{-3}   }\eqno(A.27a),$$   
$${\rm <Nz>_{Zc \nu}  \approx  1.4 \times 10^{-3}   }\eqno(A.27b),$$   
$${\rm <Nz>_{Z\tau \nu}  \approx  1.7 \times 10^{-3}   }\eqno(A.27c),$$   
where, for example, $\rm <Nz^{n}>_{Zb \nu}$ denotes the unsuppressed 
contribution coming from Z decays to b quark pairs. Comparing with the primary 
Z decays, we find that the unsuppressed primary decay contribution to 
$\rm <Nz^{2}>_{Fi}$ is about 45 times the secondary contribution, and that the 
unsuppressed primary decay contribution to $\rm <Nz>_{Fi}$ is about nine times the secondary contribution. Thus, ignoring suppression of the secondary neutrinos, we find from (A.5) that the primary and secondary decay neutrino contributions become comparable at $\rm m_{S} \approx 1.4 TeV$ for  $\rm  <Nz^{2}>_{Z\nu}$,     
 $\rm m_{S} \approx 2.2 TeV$ for  $\rm  <Nz^{2}>_{Z\overline{\nu}}$,       
$\rm m_{S} \approx 860 TeV$ for  $\rm  <Nz>_{Z\nu}$, and 
$\rm m_{S} \approx 1.3 TeV$ for  $\rm  <Nz>_{Z\overline{\nu}}$.

\newpage 


\section*{References}
[1] E.W.Kolb and M.S.Turner, The Early Universe,
(Addison-Wesley, Reading MA, 1990).
\newline[2] A.Dekel, S.M.Faber and M.Davis, in 'From the
Planck Scale to the Weak Scale', Proceedings, Santa Cruz,
California, 1987, edited by H.Haber (World Scientific,
Singapore, 1987), Vol.2.
\newline M.S.Turner, 'Dark Matter in the Universe', Fermilab
preprint FERMILAB-Conf-91/78-A.
\newline[3] J.Yang, M.S.Turner, G.Steigman, D.N.Schramm and
K.A.Olive, Astrophys.J. 281 (1984) 493
\newline K.A.Olive, D.N.Schramm, G.Steigman and T.P.Walker,
Phys.Lett. 236B (1990) 454
\newline T.P.Walker, G.Steigman, D.N.Schramm, K.A.Olive and
H.S.Kang, Astrophys.J. 376 (1991) 51. 
\newline[4] A.Guth, Phys.Rev.D23 (1981) 347.
\newline[5] G.R.Blumental, S.M.Faber, J.R.Primack and
M.Rees, Nature 311 (1984) 517.
\newline[6] G.F.Smoot et al, Astrophys.J. 396 (1992) L1;
E.L.Wright et al, ibid. 396 (1992) L13. 
\newline[7] D.Hegyi and K.A.Olive, Phys.Lett. 126B (1983)
28; Astrophys.J. 303 (1986) 56.
\newline D.Ryu, K.A.Olive and J.Silk, Astrophys.J. 353
(1990) 81. 
\newline[8] D.Richstone, A.Gould, P.Guhathakurta and C.Flynn, 
Astrophys. J. 388 (1992) 354.
\newline H.B.Richer and G.G.Fahlman, Nature 358 (1992) 383. 
\newline[9] C.Alcock et al., Nature 365 (1993) 621.
\newline E.Aubourg et al., Nature 365 (1993) 623.
\newline[10] S.P.Ahlen et al, Phys.Lett. 195B (1987) 603
\newline D.O.Caldwell et al, Phys.Rev.Lett. 61 (1988) 510;
Phys.Rev.Lett. 65 (1990) 1305.
\newline[11] D.O.Caldwell, Nucl.Phys.B (Proc.Suppl.) 31
(1983) 371.
\newline[12] W.H.Press and D.Spergel, Astrophys.J. 296 (1985) 679. 
\newline K.Greist and D.Seckel, Nucl.Phys. B283 (1987) 681; 
B296 (1988) 1034(E). 
\newline[13] S.Ritz and D.Seckel, Nucl.Phys. B304 (1988)
877.
\newline[14] A.Gould, Astrophys.J. 321 (1987) 571; 368
(1991) 610; 388 (1992) 338.
\newline[15] A.Gould, Astrophys.J. 387 (1992) 21. 
\newline[16] M.Kamionkowski, Phys.Rev. D44 (1991) 3021.
\newline[17] G.B.Gelmini, P.Gondolo and E.Roulet, Nucl.Phys.
B351 (1991) 623.
\newline[18] B.W.Lee and S.Weinberg, Phys.Rev.Lett. 39
(1977) 165.
\newline[19] V.D.Barger and R.J.N.Phillips, Collider Physics 
(Addison-Wesley, Reading MA, 1987).
\newline[20] G.W.Anderson and L.J.Hall, Phys.Rev. D42 (1992)
2685.
\newline[21] M.A.Shifman, A.I.Vainstein and V.I.Zahkarov,
Phys.Lett. 78B (1978) 443
\newline T.P.Cheng, Phys.Rev. D38 (1988) 2869
\newline H.Y.Cheng, Phys.Lett 219B (1989) 347
\newline R.Barbieri, M.Frigeni and G.F.Giudice, Nucl.Phys.
B313 (1989) 725.
\newline[22] J.Ellis and R.A.Flores, Phys.Lett. 263B (1991)
259.   
\newline[23] K.Freese, J.Frieman and A.Gould, Phys.Rev. D37
(1988) 3388.
\newline K.Greist, Phys.Rev. D38 (1988) 2357.
\newline[24] P.F.Smith and J.D.Lewin, Phys.Rep. 187 (1990)
203.
\newline P.F.Smith et al, Phys.Lett. 255B (1991) 451.
\newline [25] B.Cabrera, Proceedings of the IFT Conference
on Dark Matter, Gainsville, Florida (1992).
\newline J.Ellis and R.A.Flores, CERN-TH-6483/92 (1992)
\newline[26] IMB Collaboration, J.M.LoSecco et al,
Phys.Lett. 188B (1987) 388.
\newline R.Svoboda et al, Astrophys.J. 315 (1987) 420.
\newline[27] F.Halzen, T.Stelzer and M.Kamionkowski,
Phys.Rev. D45 (1992) 4439.
\newline[28] R.C.Webb, in Proceedings of the International
School of Astroparticle Physics, Ed. D.V.Nanopoulos, World
Scientific, Singapore (1993).
\newline[29] A.Okada et al, Proceedings of the Workshop on
High Energy Neutrino Astrophysics, edited by V.J.Stenger et
al, World Scientific, Singapore (1992)
\newline DUMAND Collaboration, Phys.Rev. D42 (1990) 3613.
\newline[30] S.Tilev et al, Proceedings of the 23rd
International Cosmic Ray Conference, Calgary, Canada (1993).
\newline[31] Proceedings of the NESTOR Workshop, edited by
L.K.Resvanis, Univ. of Athens (1993).
\newline[32] F.Halzen and J.G.Learned, Univ. of
Wisconsin-Madison preprint MAD/PH/759 (1993).
\newline J.G.Learned, Nucl.Phys.B (Proc.Suppl.) 31 (1993)
456.
\newline[33] Frejus Collaboration, presented by H.J.Daum,
Topical Seminar on Astrophysics and Particle Physics, San
Miniato, Italy 1989.

\section*{Figure Captions}
{\bf Figure 1a} S annihilation to $\rm h^{o}$
pairs.
\newline {\bf Figure 1b} S annihilation to W and Z pairs.
\newline {\bf Figure 1c} S annihilation to fermion pairs.
\newline {\bf Figure 2a} Thermal relic S scalar density
(in units of $\rm \Omega_{S}h^{2}$) as a function of $\rm
\lambda_{S}$ and $\rm m_{S}$ (in units of GeV) for $\rm
m_{h}$ equal to 60GeV.
 \newline {\bf Figure 2b} Thermal relic S scalar density  
for $\rm m_{h}$ equal to 100GeV.
 \newline {\bf Figure 2c} Thermal relic S scalar density
for $\rm m_{h}$ equal to 300GeV.
 \newline {\bf Figure 2d} Thermal relic S scalar density
for $\rm m_{h}$ equal to 500GeV.
\newline {\bf Figure 3a} Ge scattering rate for the case
$\rm m_{h} = 60GeV$. The contours correspond to $\rm
\sigma_{S-Ge} = $ a) $\rm 1000kg^{-1}d^{-1}$, b) $\rm
100kg^{-1}d^{-1}$ , c) $\rm 10kg^{-1}d^{-1}$, d) $\rm
1kg^{-1}d^{-1}$ , e) $\rm 0.1kg^{-1}d^{-1}$ and f) $\rm
0.01kg^{-1}d^{-1}$. 
 \newline {\bf Figure 3b} Ge scattering rate for the case 
$\rm m_{h} = 100GeV$.
 \newline {\bf Figure 3c} Ge scattering rate for the case 
$\rm m_{h} = 300GeV$.
\newline {\bf Figure 4a} Rate of upward-moving muons at the
Earth's surface due to neutrinos from S annihilation in the
Sun, for the case $\rm m_{h} = 60GeV$. The contours
correspond to $\rm \Gamma_{detector} = $
 a) $\rm 2.65x10^{-2}m^{-2}yr^{-1}$, b) $\rm
10^{-3}m^{-2}yr^{-1}$, c) $\rm 10^{-4}m^{-2}yr^{-1}$, d)
$\rm 10^{-5}m^{-2}yr^{-1}$ and e) $\rm
10^{-6}m^{-2}yr^{-1}$.
\newline {\bf Figure 4b} Rate of upward-moving muons at the
Earth's surface due to neutrinos from S annihilation in the
Sun, for the case $\rm m_{h} = 100GeV$.
\newline {\bf Figure 4c} Rate of upward-moving muons at the
Earth's surface due to neutrinos from S annihilation in the
Sun, for the case $\rm m_{h} = 300GeV$.
\newline {\bf Figure 5a} Rate of upward-moving muons at the
Earth's surface due to neutrinos from S annihilation in the
Earth, for the case $\rm m_{h} = 60GeV$. The contours
correspond to $\rm \Gamma_{detector} = $
 a) $\rm 2.65x10^{-2}m^{-2}yr^{-1}$, b) $\rm
10^{-3}m^{-2}yr^{-1}$, c) $\rm 10^{-4}m^{-2}yr^{-1}$, d)
$\rm 10^{-6}m^{-2}yr^{-1}$.
\newline {\bf Figure 5b} Rate of upward-moving muons at the
Earth's surface due to neutrinos from S annihilation in the
Earth, for the case $\rm m_{h} = 100GeV$.
\newline {\bf Figure 5c} Rate of upward-moving muons at the
Earth's surface due to neutrinos from S annihilation in the
Earth, for the case $\rm m_{h} = 300GeV$.
\newline {\bf Figure 6a} Rate of contained events due to
neutrinos from S annihilation in the Sun, for the case $\rm
m_{h} = 60GeV$. The contours correspond to $\rm
\Gamma_{detector} = $
 a) $\rm 6.4kt^{-1}yr^{-1}$, b) $\rm 1kt^{-1}yr^{-1}$, c)
$\rm 0.1kt^{-1}yr^{-1}$, d) $\rm 10^{-2}kt^{-1}yr^{-1}$ and
e) $\rm 10^{-3}kt^{-1}yr^{-1}$.
\newline {\bf Figure 6b} Rate of contained events due to
neutrinos from S annihilation in the Sun, for the case $\rm
m_{h} = 100GeV$.
\newline {\bf Figure 7a} Rate of contained events due to
neutrinos from S annihilation in the Earth, for the case
$\rm m_{h} = 60GeV$. The contours correspond to $\rm
\Gamma_{detector} = $
 a) $\rm 6.4kt^{-1}yr^{-1}$, b) $\rm 1kt^{-1}yr^{-1}$, c)
$\rm 0.1kt^{-1}yr^{-1}$, d) $\rm 10^{-2}kt^{-1}yr^{-1}$ and
e) $\rm 10^{-3}kt^{-1}yr^{-1}$.
\newline {\bf Figure 7b} Rate of contained events due to
neutrinos from S annihilation in the Earth, for the case
$\rm m_{h} = 100GeV$.

\newpage

\begin{figure}[h] 
                    \centering                   
                    \includegraphics[width=0.8\textwidth, angle=0]{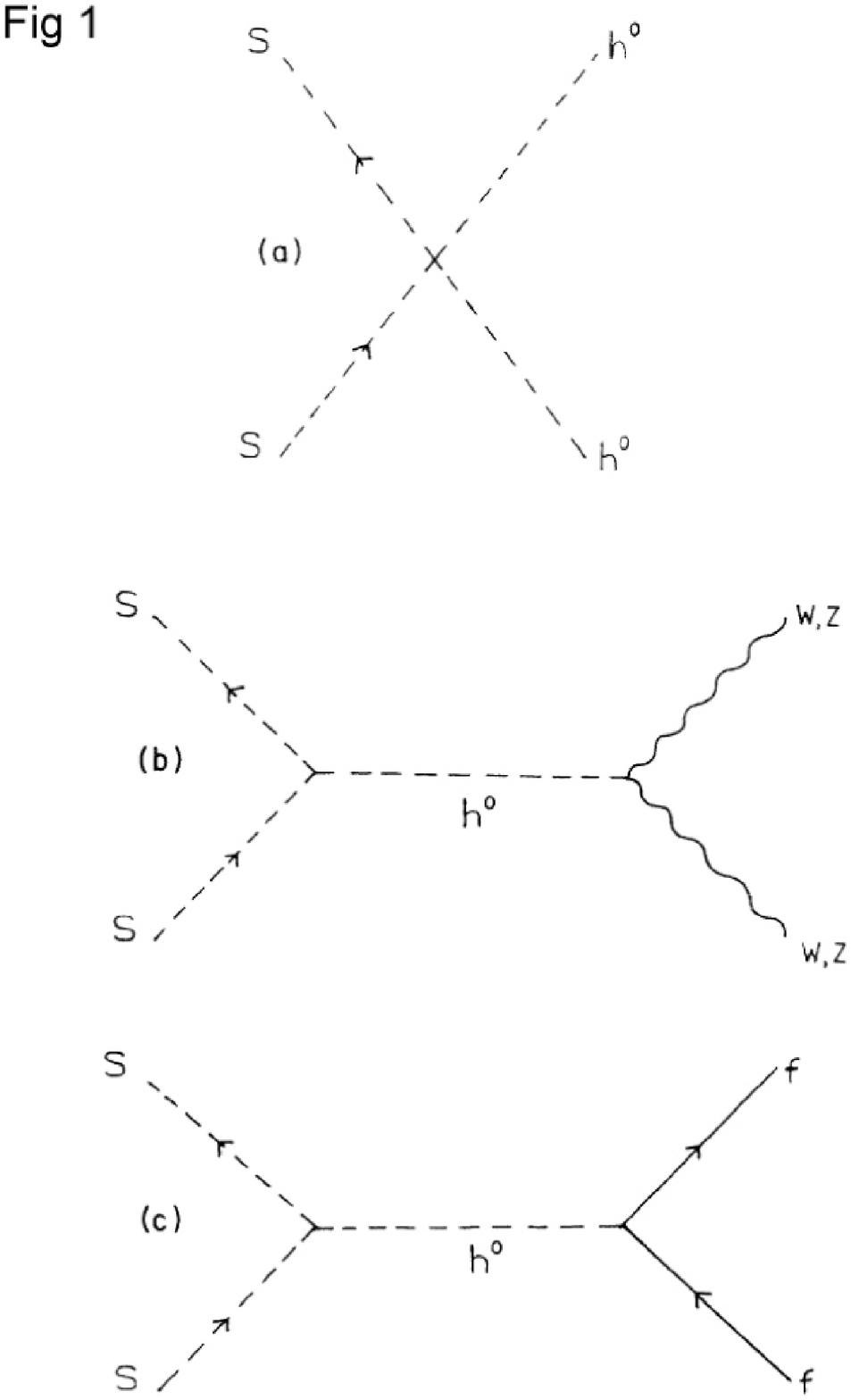}
                    
                    \end{figure}

\newpage

\begin{figure}[h] 
                    \centering                   
                    \includegraphics[width=0.7\textwidth, angle=0]{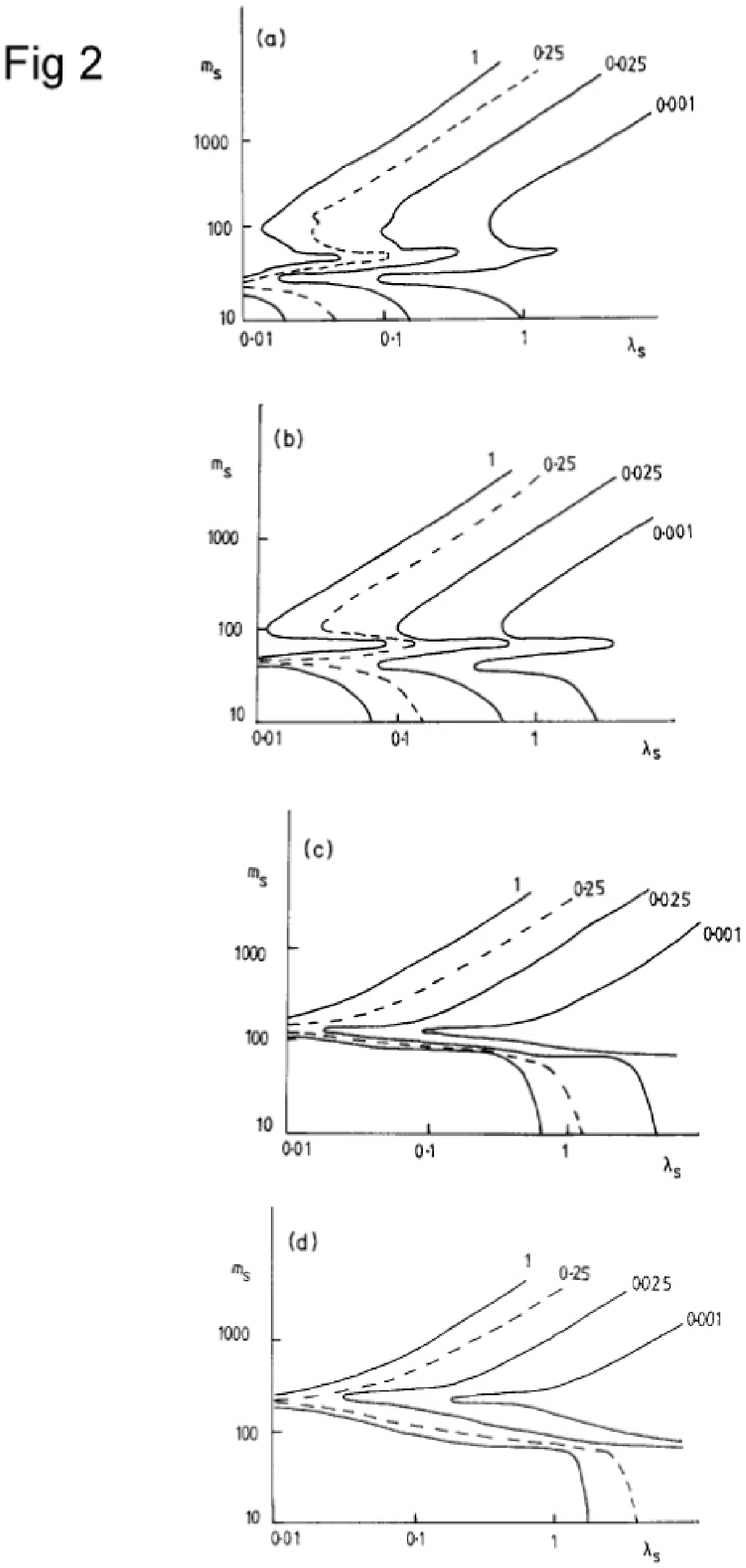}
                    
                    \end{figure}

\newpage

\begin{figure}[h] 
                    \centering                   
                    \includegraphics[width=0.8\textwidth, angle=0]{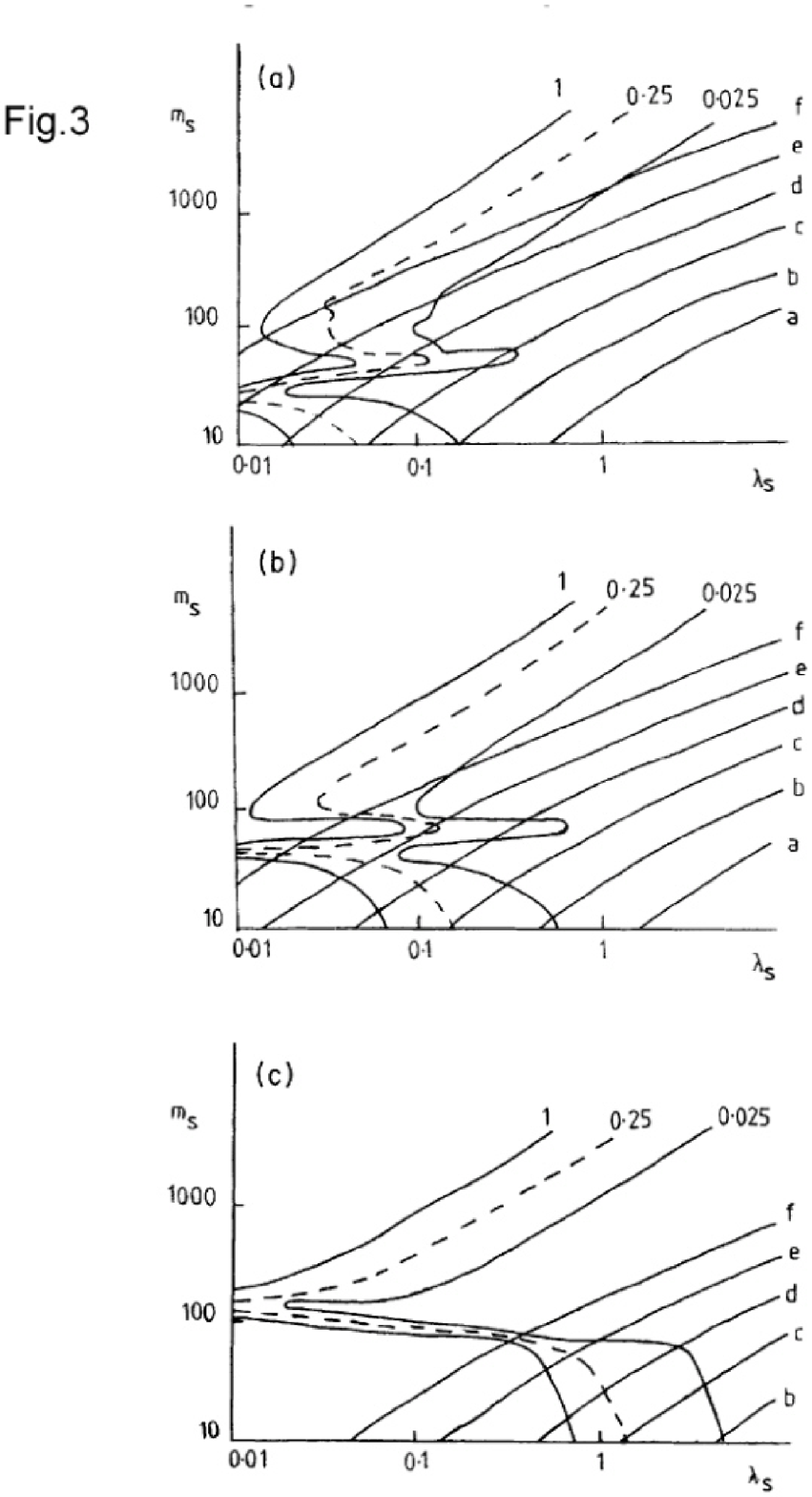}
                    
                    \end{figure}

\newpage

\begin{figure}[h] 
                    \centering                   
                    \includegraphics[width=0.8\textwidth, angle=0]{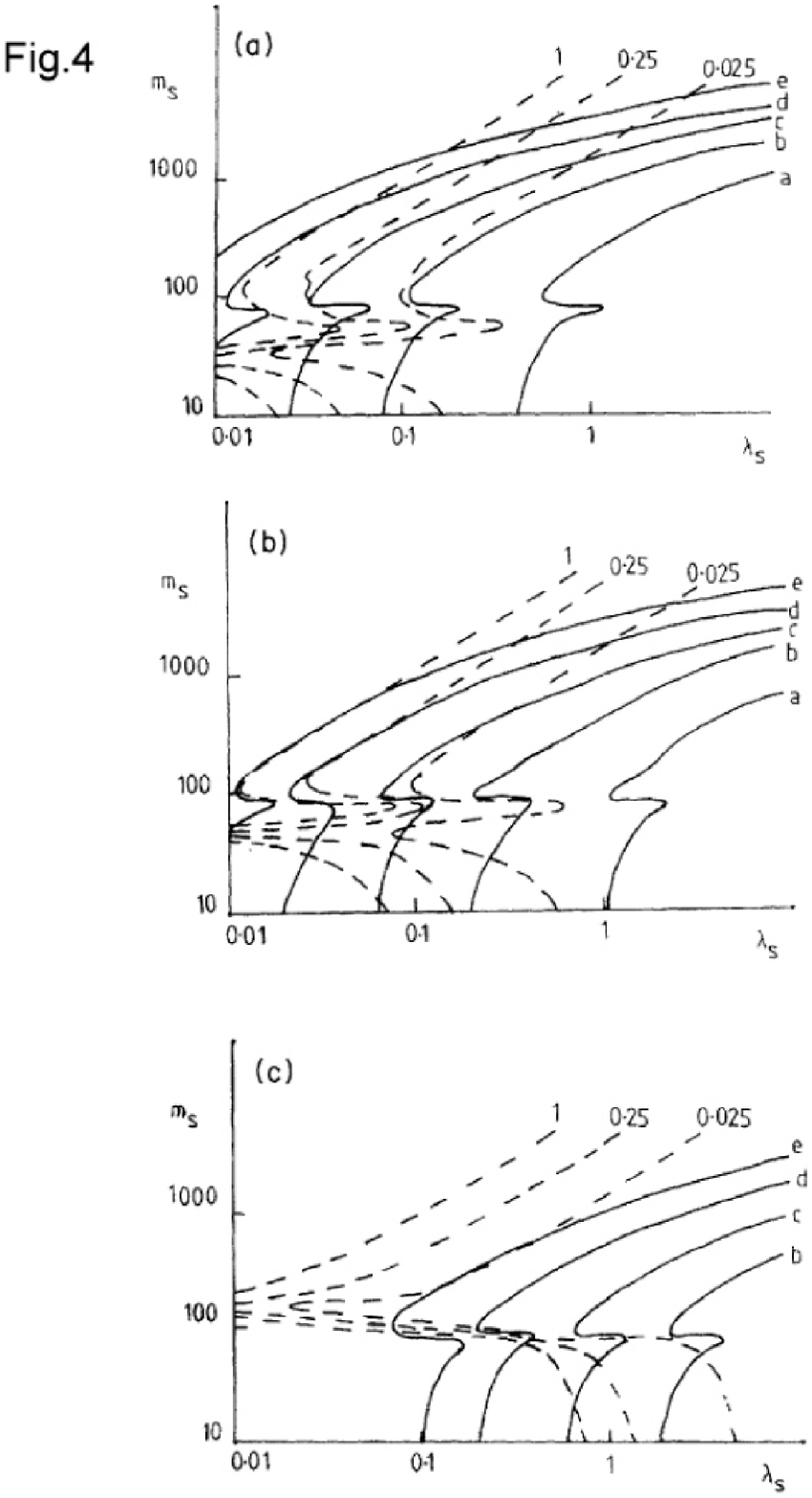}
                    
                    \end{figure}

\newpage

\begin{figure}[h] 
                    \centering                   
                    \includegraphics[width=0.8\textwidth, angle=0]{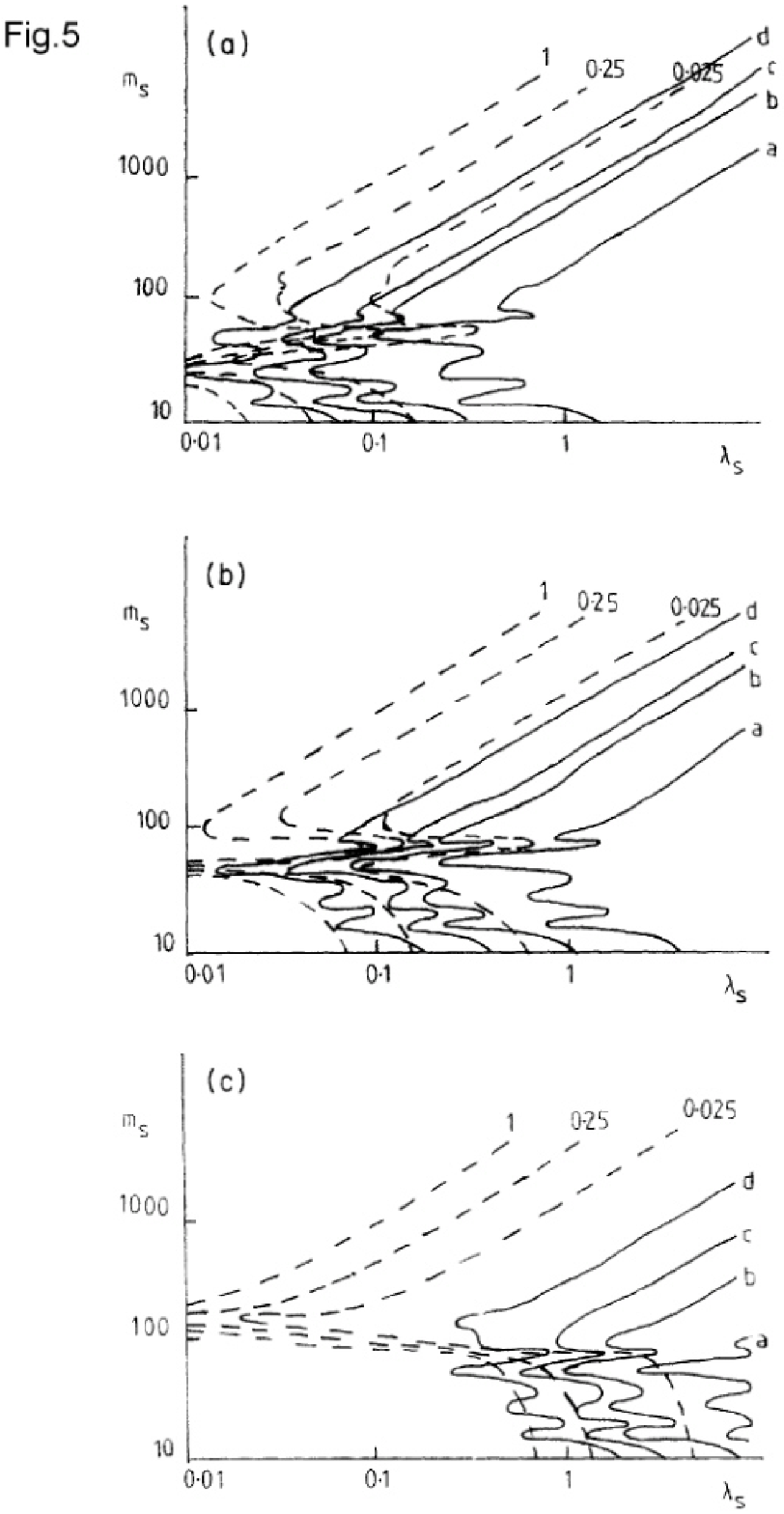}
                    
                    \end{figure}

\newpage

\begin{figure}[h] 
                    \centering                   
                    \includegraphics[width=1.0\textwidth, angle=0]{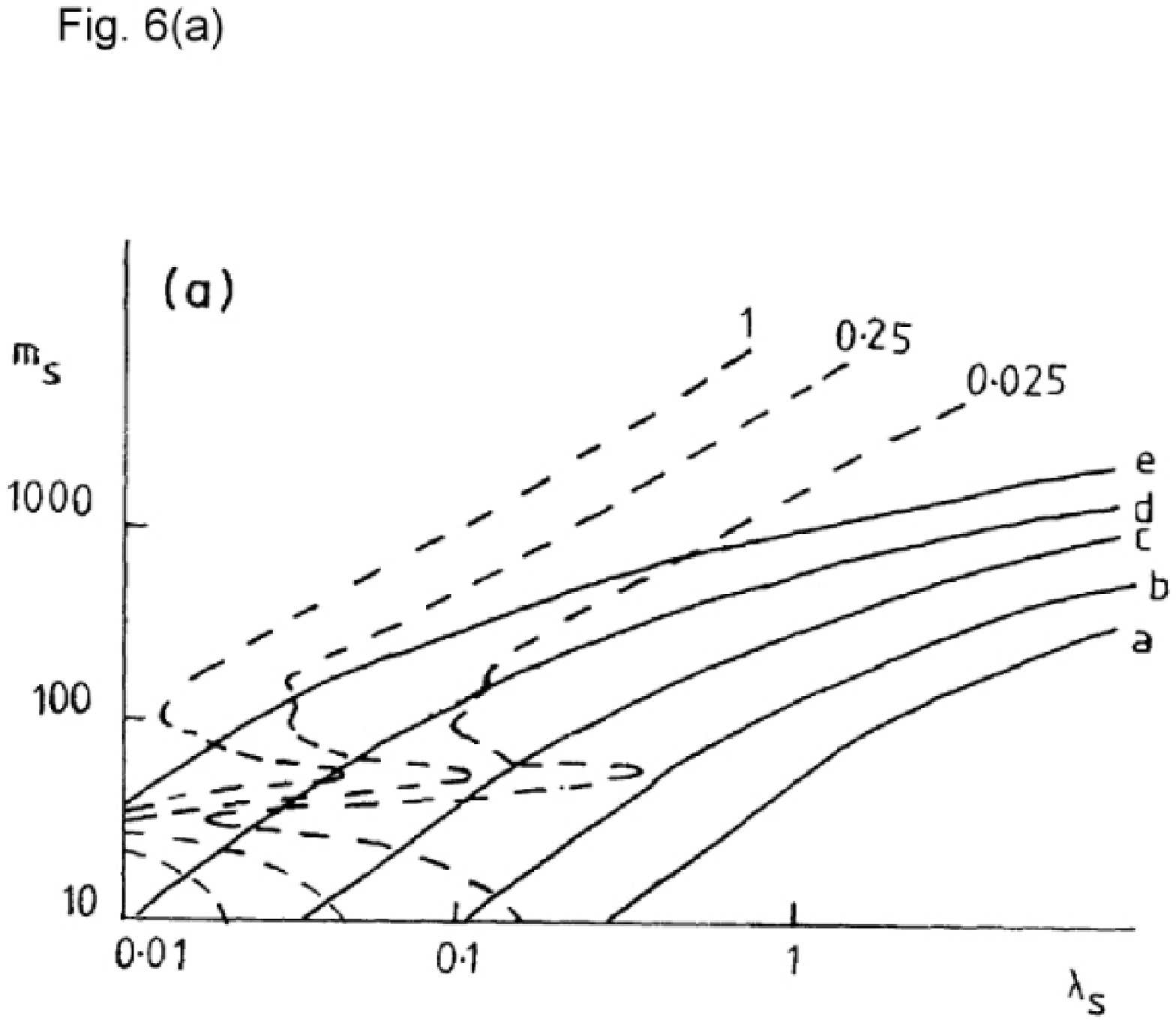}
                    
                    \end{figure}

\newpage

\begin{figure}[h] 
                    \centering                   
                    \includegraphics[width=1.0\textwidth, angle=0]{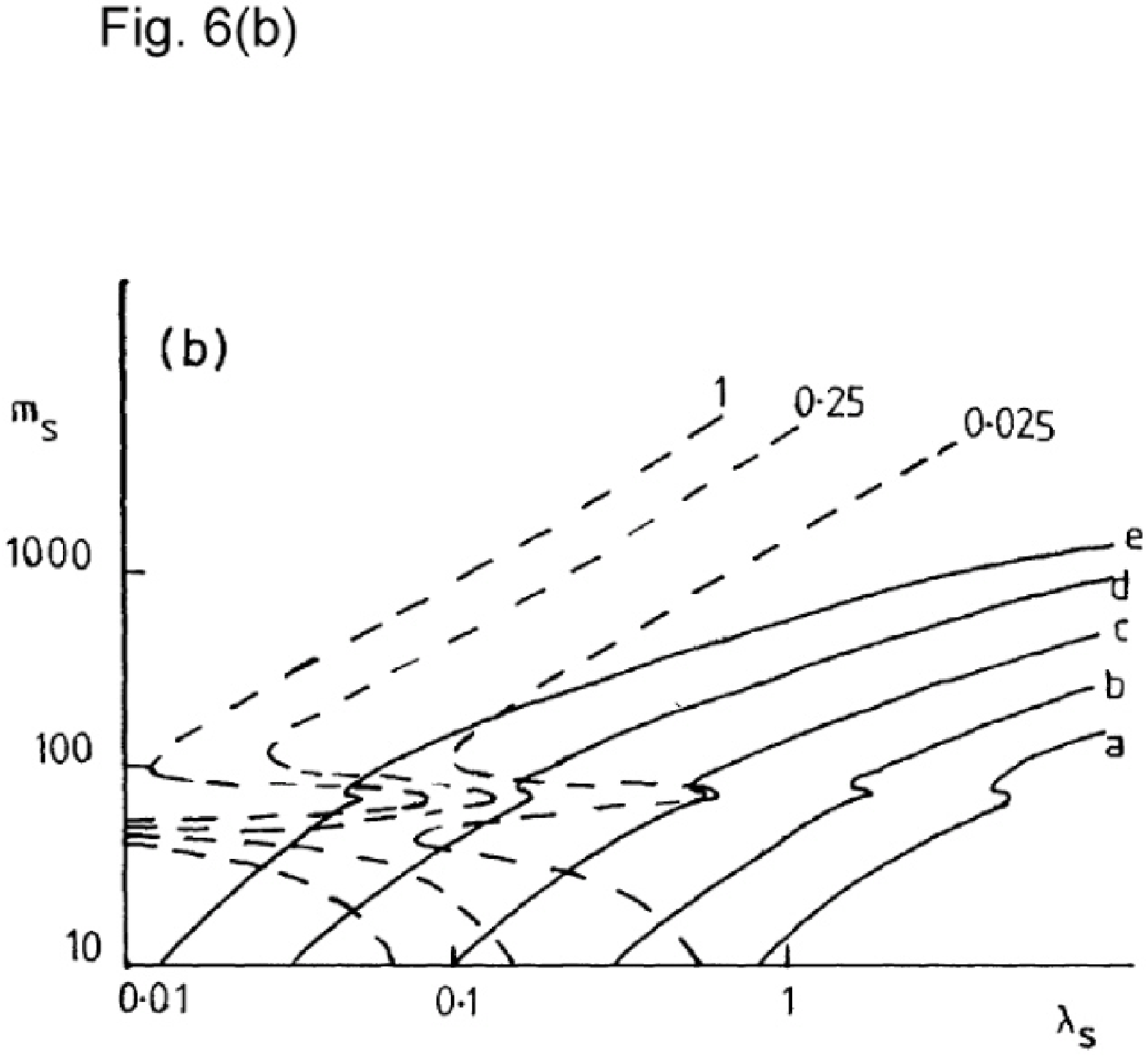}
                    
                    \end{figure}

\newpage

\begin{figure}[h] 
                    \centering                   
                    \includegraphics[width=1.0\textwidth, angle=0]{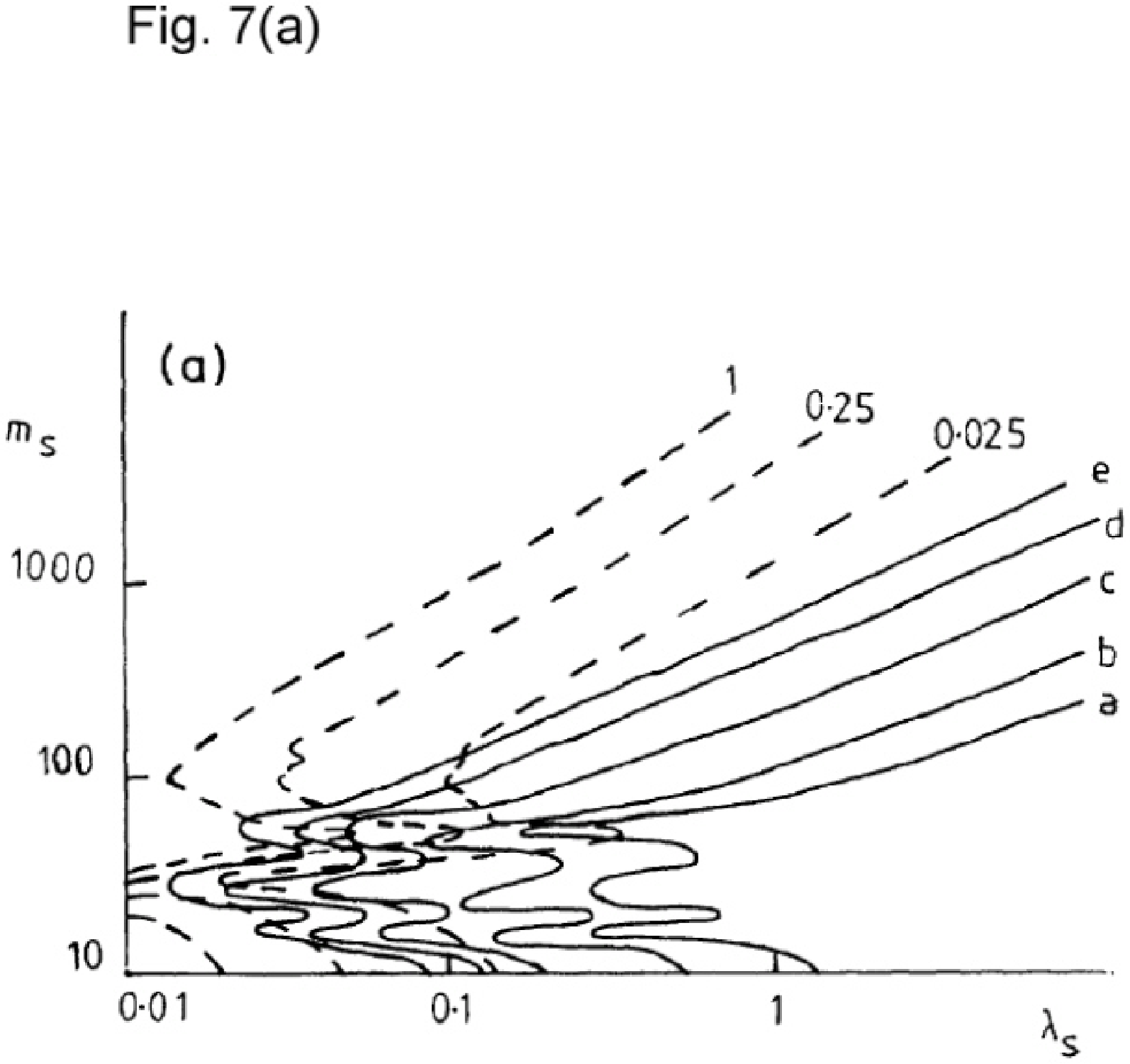}
                    
                    \end{figure}   

\newpage

\begin{figure}[h] 
                    \centering                   
                    \includegraphics[width=1.0\textwidth, angle=0]{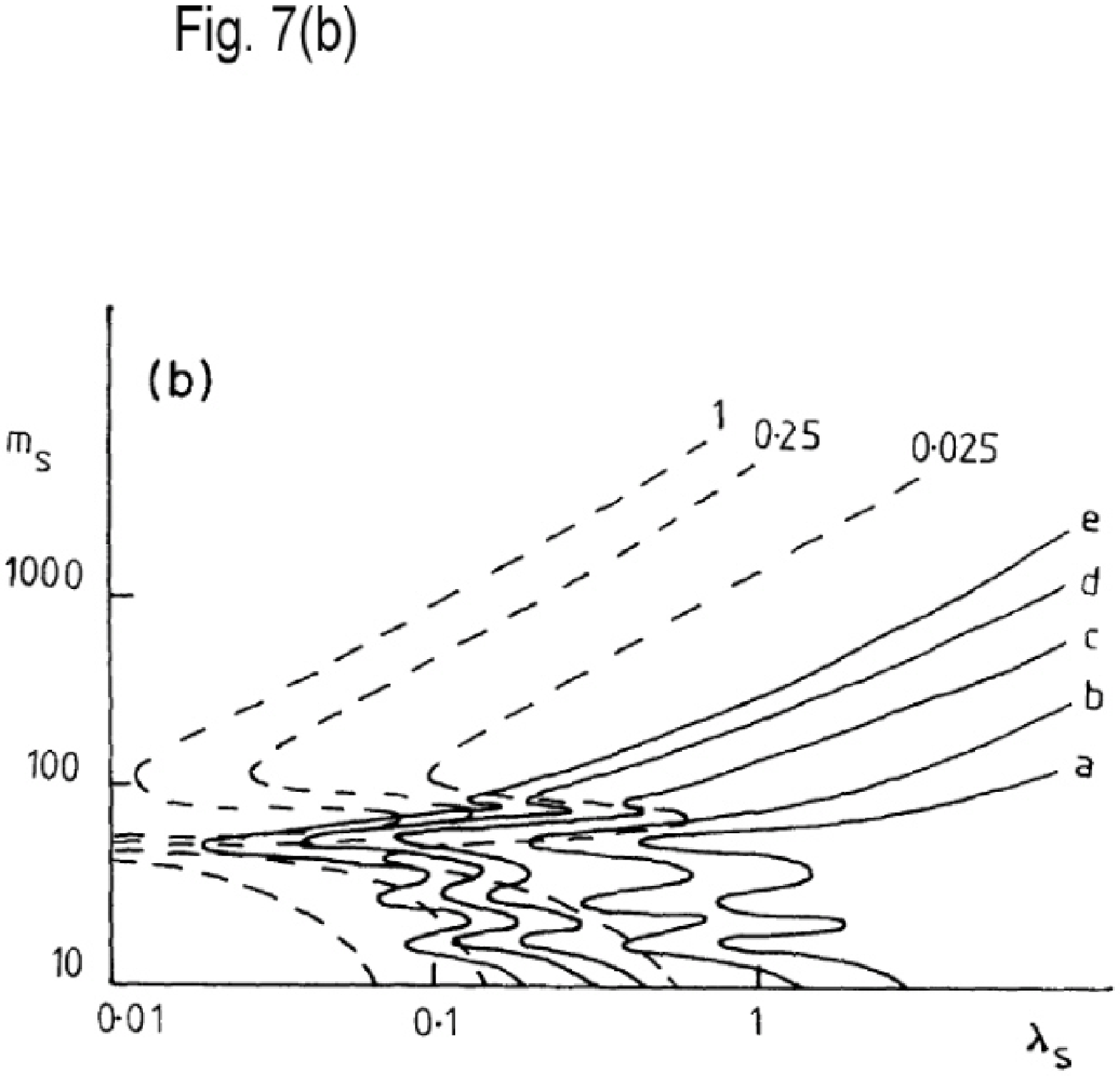}
                    
                    \end{figure}

\end{document}